\documentclass[11pt,oneside,fleqn,reqno]{article}
\usepackage{amsmath}
\usepackage{amsthm}
\usepackage{amssymb}
\usepackage[pdftex]{graphicx,color}
\usepackage{mathtools}
\usepackage{wrapfig}
\usepackage{natbib} \citeindextrue
\usepackage{subfigure}
\usepackage{dsfont}
\usepackage{multirow}
\usepackage{multicol}
\usepackage{algorithm}
\usepackage{algorithmic}
\usepackage{bm}
\usepackage{comment}
\usepackage{url}

\usepackage{authblk}

\DeclareMathOperator{\tr}{tr}

\allowdisplaybreaks

\def \qed{\hfill $\Box$}

\def \notin{{\not \in}}

\def \hbar{\bar h}

\def \v1 {\mathbb{F}}

\newcommand{\doublespace}{\addtolength{\baselineskip}{.5\baselineskip}}
\newcounter{stepno}

\newtheorem{thm}{Theorem}[section]

\newtheorem{lem}{Lemma}[section]

\newtheorem{prop}{Proposition}[section]

\newcommand{\bnn}{\\ \\$\begin{array}{rcll}}
\newcommand{\enn}{\end{array}$\\ \\}
\newcommand{\ba}{\begin{eqnarray}}
\newcommand{\ea}{\end{eqnarray}}
\newcommand{\bas}{\begin{eqnarray*}}
\newcommand{\eas}{\end{eqnarray*}}
\newcommand{\bdm}{\begin{displaymath}}
\newcommand{\edm}{\end{displaymath}}
\newcommand{\be}{\begin{equation}}
\newcommand{\ee}{\end{equation}}
\newcommand{\bn}{\begin{eqnarray}}
\newcommand{\en}{\end{eqnarray}}
\newcommand{\bns}{\begin{eqnarray*}}
\newcommand{\ens}{\end{eqnarray*}}

\newcommand{\defvarbegin}{\begin{quotation}\vspace{-15pt}\begin{tabbing}}
\newcommand{\defvarend}  {\end{tabbing}\vspace{-10pt}\end{quotation}}

\newcommand{\bnarray}{\begin{equation}\begin{array}{rcll}}
\newcommand{\enarray}{\end{array}\end{equation}}

\newcommand{\barr}{\begin{array}}
\newcommand{\earr}{\end{array}}

\newcounter{cnum}

\newcommand{\beginalg}{\setcounter{stepno}{1}
                \begin{list}{\bf Step~\arabic{stepno}}
                         {\usecounter{stepno}\settowidth{\labelwidth}{\bf Step~9m}
                \addtolength{\leftmargin}{2\parindent}}
                }
\newcommand{\eg}{\end{list}}

\def \define{\begin{quote}\begin{itemize}}
\def \enddefine{\end{itemize}\end{quote}}

\newlength{\boxedparwidth} 
  {\begin{center} \begin{tabular}{|@{\hspace{.15in}}c@{\hspace{.15in}}|}
                 \hline \\ \begin{minipage}[t]{\boxedparwidth}
                 \setlength{\parindent}{0.0in}}%
   {\end{minipage} \\ \\ \hline \end{tabular} \end{center}}
\newcounter{chapterctr}
\newcounter{example}[chapterctr]
\newcounter{ctr}
\newtheorem{assum}{Assumption}[section]

\begin{document}

\title{Optimal data-driven hiring with equity for underrepresented groups}

\author[1]{Yinchu Zhu \thanks{yinchuzhu@brandeis.edu}}
\author[2]{Ilya O. Ryzhov \thanks{iryzhov@umd.edu}}
\affil[1]{Department of Economics, Brandeis University}
\affil[2]{Robert H. Smith School of Business, University of Maryland}

\date{\today}

\maketitle

\begin{abstract}
We present a data-driven prescriptive framework for fair decisions, motivated by hiring. An employer evaluates a set of applicants based on their observable attributes. The goal is to hire the best candidates while avoiding bias with regard to a certain protected attribute. Simply ignoring the protected attribute will not eliminate bias due to correlations in the data. We present a hiring policy that depends on the protected attribute functionally, but not statistically, and we prove that, among all possible fair policies, ours is optimal with respect to the firm's objective. We test our approach on both synthetic and real data, and find that it shows great practical potential to improve equity for underrepresented and historically marginalized groups.
\end{abstract}

\doublespace

\section{Introduction}\label{sec:intro}

We have just received an application for a job, or perhaps for admission to an academic program. From the application, we obtain a vector $X\in\mathds{R}^p$ describing the qualifications of the applicant -- educational background, years of work experience, GPA, test scores, and other information relevant to our decision. Suppose that the performance of the applicant, if hired or admitted, can be quantified and predicted using the linear regression model
\begin{equation}\label{eq:reg}
Y = \beta^\top X + \varepsilon,
\end{equation}
where $\beta\in\mathds{R}^p$ is a vector of fixed effects, and $\varepsilon$ is an independent, zero-mean noise term. Although $\beta$ is unknown, let us suppose that it can be estimated using the observed performance of past applicants. Given an estimator $\hat{\beta}$, we can calculate the predicted performance $\hat{\beta}^\top X$ and use this to make our decision.

The situation becomes more complicated when, in addition to $X$, the application provides an additional \textit{protected attribute} $Z\in\left\{0,1\right\}$ which indicates whether or not the applicant belongs to a certain disadvantaged demographic or socioeconomic group. Many such attributes are explicitly indicated in job or academic applications; in other cases, the protected information can be obtained with extremely high accuracy from the applicant's publicly observable social media presence \citep{KoStGr13}. Either way, let us suppose that this information is known to the decision-maker. Now, for ethical and legal reasons, we must ensure that the hiring or admissions decision is fair with respect to the protected attribute.

However, the word ``fair'' has many meanings. The decision is based on model (\ref{eq:reg}), in which $Z$ is not present; in other words, $Z$ has no \textit{direct} or \textit{causal} effect on the performance $Y$. Thus, two applicants with different values of $Z$, but identical values of $X$, will have the same expected performance under the model and will be ranked identically. One could therefore argue that such an approach, where the ranking has no \textit{functional} dependence on $Z$, is fair. This concept, known as ``parity of treatment,'' continues to be one of the most prominent criteria used to assess fairness (see Section \ref{sec:litreview} for a detailed discussion).

The problem is that \textit{functional} independence does not ensure \textit{probabilistic} independence: there may be correlations in the joint (``data-generating'') distribution of $\left(X,Z\right)$, arising from causal factors that are not observable to the decision-maker. For example, evidence indicates that test scores are influenced by structural bias in education \citep{PeRuOsSi16}, in the design of the tests themselves \citep{SaWi10}, and in unequal access to opportunity \citep{LuSa20}, but to the decision-maker in our motivating example, all these factors will show up only as a negative correlation between test scores and membership in a disadvantaged group. Consequently, the random variable $\beta^\top X$ will also be correlated with $Z$, and any decision based solely on expected performance will systematically disfavor applicants who belong to that group. Thus, parity of treatment will end up reinforcing inequality, a problem well-known to legal \citep{Ha00} and moral \citep{Th15} philosophy. In this paper, we show analytically how this occurs: when the candidate pool is large, the hiring decision is determined by the extreme values of $\beta^\top X$, which has the effect of amplifying even seemingly minor disparities.


Various solutions have been debated, e.g., reducing reliance on test scores \citep{SpRa19}, or redacting the names of applicants from r\'{e}sum\'{e}s to prevent discrimination \citep{BeMu04}. However, experimental evidence suggests that such policies may have unintended adverse impact \citep{BeCrBa15}, while meeting with resistance among hiring managers \citep{FoWi18} who interpret them as a sign that the organization does not trust them. In an effort to approach disparities more systematically, firms are turning to AI and machine learning tools to guide hiring and other decisions (see \citealp{Finocchiaro21} for a high-level overview). Such tools are likely to become more prevalent with time: \cite{RaBaKlLe20} discusses the legal considerations and argues that ``the ready availability of algorithmic techniques might effectively create a legal \textit{imperative} to use them.'' Practical results, however, have not always lived up to expectations \citep{SuHiLa21}. In one widely publicized instance, an AI tool at Amazon was found to favor certain features of r\'{e}sum\'{e}s that were more likely to appear on applications submitted by men \citep{BlEs20}, a clear example of the pitfall of probabilistic dependence. In our view, these issues demonstrate the need for a transparent prescriptive framework that does not withhold information from decision-makers, but rather uses it fairly while also provably meeting firms' objectives (such as hiring the best candidates).

This paper proposes such a framework. We use a stylized model of the hiring problem, where (\ref{eq:reg}) is used to predict performance, and the goal is to select a candidate from a pool that is drawn randomly from a population. We create a hiring policy that depends on $Z$ \textit{functionally}, but is independent of $Z$ \textit{probabilistically} -- far from ``redacting'' protected attributes, it is necessary to use them in order to ensure fairness. This policy is not only fair, but also \textit{optimal} with respect to the employer's objective, and it is transparent and interpretable: the employer can see exactly how performance and fairness are considered. Our approach proceeds in three parts:

\begin{enumerate}
\item We first derive an \textit{ideal fair policy} that assumes knowledge of both $\beta$ and the joint distribution of $\left(X,Z\right)$. The decision directly compares the best-performing candidates from each subgroup and determines a performance threshold for hiring the minority candidate. This threshold is both fair and optimal in the sense of maximizing performance.
\item When the problem inputs are unknown, but can be estimated from historical data, we prove that there is no non-trivial ranking that can achieve perfect fairness in this setting. There will always be deviation from the ideal fair policy; the question is how much.
\item We show how the ideal policy can be estimated in a manner that is not perfectly fair, but \textit{optimal} in the sense of converging to the ideal policy at the fastest possible rate (we give matching upper and lower bounds) with respect to the size of the available data.
\end{enumerate}

A key insight of our work is that the presence of an explicit decision problem (optimizing over the candidate pool) has a profound and surprising impact on the structure of the ideal fair policy. As explained in Section \ref{sec:litreview}, recent work has studied optimal fair \textit{prediction} of expected performance, i.e., finding the closest approximation of $\beta^\top X$ that is independent of $Z$. We prove, however, that ranking candidates by optimal fair predictions of their performance is \textit{not} optimal in the hiring problem, because the decision requires us to compare across subgroups as well as within them. This result strongly distinguishes our work from the rich literature on fair prediction.

Our theoretical results are complemented by numerical experiments on both synthetic (simulated) data as well as real data describing over $20,000$ law students. In these experiments, the incidence of the protected group ranges from as low as $6\%$ to as high as $43\%$. In all cases, our method scales well, is easy to implement, and significantly outperforms alternative approaches in reducing discrimination. In many practical situations, fairness is achievable with negligible performance reduction: in other words, it is possible to uphold employers' interests while also reducing harm to minority candidates.


The OR/OM community has become increasingly aware of diversity, equity, and inclusion as a subject for research \citep{JoCh20}. Recent surveys and tutorials, such as \cite{FuHuSi20} and \cite{BjAn20}, draw heavily from the machine learning community, where fairness has been a subject of active research. Our work uses the stylized hiring problem to rigorously demonstrate the crucial and subtle distinction that arises when fairness is viewed from an \textit{operations} perspective, i.e., in the context of a decision problem, where the goal is to find an optimal solution and not only to improve predictive power.

\section{Related work}\label{sec:litreview}

In the OR/OM literature, ``fairness'' has generally referred to equitable outcomes across a given set of entities \citep{BeFaTr11}, e.g., retailers in a supply chain game \citep{HaRaZh07}, locations served by a doctor \citep{McLe14,Rea21}, or clients of an investment firm \citep{Diana21}. Here, fairness is viewed in a game-theoretic, rather than statistical sense; the set of entities is pre-specified, not sampled from a population, and data collection is usually not explicitly modeled. On the other hand, there is also a growing body of empirical/experimental literature on bias and discrimination in, e.g., the sharing economy \citep{CuLiZh20,MePa21}, as well as on algorithmic bias \citep{LaTu19}. This work complements ours, by showing that the problem exists, but it generally does not propose analytical methods to alleviate it.

Our work is closer to the literature on prescriptive analytics and data-driven decision-making \citep{MiPe20}, which uses statistical methods ranging from linear regression \citep{BeOhReSi16} to classification \citep{ChWaWa18} and machine learning \citep{BeKa20} to learn optimal decisions \citep{Mi20} from observational data \citep{HoLiWa18}. This work sees prediction, not as an end in itself, but rather as a means for improved decision-making \citep{ElGr21}. Data-driven methods have provided new insight into classical optimization problems such as the newsvendor \citep{LePeUi15,BaRu19}. However, this active area of research has not yet explored fairness. Conversely, recent work by \cite{SaGu19} on fair decisions does not model data (regression features describing individual observations).

Many, perhaps most, existing data-driven models of fairness originate from the computer science and machine learning communities. Although there are hundreds of references on ``algorithmic bias'' or ``fair prediction'' (see \citealp{BaGoLo20} for one survey), the precise definition of ``fairness'' varies greatly from one group of papers to the next. There are inherent, unavoidable tradeoffs between these types of fairness \citep{KlMuRa17}. It is useful to identify four broad categories of papers and assess each respective definition of fairness in the context of data-driven hiring:

\textit{1) ``Fairness'' as ``making similar decisions for similar individuals.''} Studies in this category have considered prediction \citep{HeKr18} as well as decision-making \citep{GuKa21}, with one specific context being auctions \citep{IlJaCh20,ChJa21}. The unifying theme of these papers is that two individuals with the same attributes should receive the same treatment (or as close to it as possible). This issue arises in online decision-making \citep{GiJuKeRo18} where one wishes to avoid penalizing candidates who appeared early on, when the data were not sufficient to accurately assess them.

We do not consider this notion of fairness because it cannot help disadvantaged groups. In fact, a policy that consistently discriminates against certain values of $Z$ technically does treat \textit{similar} candidates in the same way. Forcing such a policy to ignore $Z$ will not help either, because $Z$ is correlated with $X$, so candidates with similar values of $X$ are more likely to have the same $Z$ as well. Our goal in this work is to ensure a measure of equity for candidates who are \textit{different}.

\textit{2) ``Fairness'' as parity of treatment.} Parity of treatment \citep{BaSe16} means that our decision is conditionally independent of $Z$ given $X$. This notion was considered in such decision-making contexts as bandit problems \citep{JoKeMoRo16}, reinforcement learning \citep{Jabbari17}, and subset selection \citep{KeRoWu17}. It is sometimes called ``meritocratic'' fairness; in the specific context of hiring, it was succinctly described by \cite{JoKeMoRo16} as, ``a worse applicant is never favored over a better one.''

Unfortunately, the situation is not so simple. When an applicant is perceived to be ``worse'' because of structural bias, this approach to decision-making simply reinforces that bias. What is more, as shown in Section \ref{sec:noparity} of our paper, the disparity between groups is \textit{amplified} by the competitiveness of the position: even if this disparity is small on an \textit{individual} level, a ``meritocratic'' policy will end up completely excluding the disadvantaged group when the number of applicants is large. In our view, this poses a serious ethical problem.

\textit{3) ``Fairness'' as parity of error.} We use the term ``parity of error'' to collectively refer to several related notions such as equality of opportunity \citep{HaPrPrSr16}, equality of chances \citep{LoHeHe21}, and predictive parity \citep{Ch17}. Studies in this category primarily focus on building a \textit{prediction} $\hat{Y}$ of $Y$, though, e.g., \cite{Blum18} has applied this concept to decision-making. Their main goal is to eliminate bias in the estimation error: for example, in risk adjustments in healthcare \citep{ZiRo20}, predictions may systematically under- or over-estimate costs for various patient classes, and it is necessary to balance the estimation error between them. In classification problems, where the response $Y$ is binary, this can be expressed simply as balancing false positives and/or negatives across different $Z$ values \citep{CaVe10,GoBaFaLo19,BlLy20}.

Parity of error can be important when each individual is considered in isolation, e.g., the decision to offer a loan, or to grant parole, is based only on that particular person's predicted risk. \cite{Feldman15} uses hiring to motivate parity of error, but interprets it as a yes/no decision made for each individual separately, thus reducing the decision problem to a prediction problem. In actuality, however, the hiring decision selects \textit{between} candidates, and the rule for making this selection is a higher-priority issue than estimation. If $\beta$ in (\ref{eq:reg}) is known, there is no estimation error at all, but using $\beta^\top X$ to rank candidates is still unfair for the reasons discussed previously. Our numerical experiments show that existing parity of error methods are unable to ensure representation for disadvantaged groups.


\textit{4)``Fairness'' as statistical parity.} Statistical parity means that our decision is independent of $Z$. This is the criterion adopted in our paper. It has been criticized in the literature on the grounds that it can be easily gamed \citep{Dwork2012,KeNeRoWu18}, for example by hiring at random from the protected group, or through other manipulations that reduce population-wide measures of disparity (see \citealp{AnBjDeRo19} for some examples in the context of the gender pay gap). In our setting, however, these issues are overcome by the presence of a decision problem with an economic objective function: among all policies that are fair in this sense, we wish to find the \textit{best-performing} one with respect to the employer's interests.

The literature on promoting diversity in algorithmic decision-making, in contexts such as ranking \citep{CeStVi18,Zehlike18} and admissions \citep{Borgs19}, has a similar goal of ensuring fair representation of disadvantaged groups. However, these papers take a ``deterministic'' approach to the problem by simply placing hard lower bounds on the proportion of candidates to be selected from the protected subgroup, regardless of their performance. By contrast, our decision is based on both the number and performance of candidates in each subgroup, and we only hire candidates who meet a (fair) performance threshold.

Among those papers that directly adopt statistical parity as the fairness criterion, many focus on prediction, and find it convenient to turn the fairness condition into a regularization penalty that can be added to any standard statistical model (see, e.g., \citealp{Zemel13}). This makes fairness into a soft constraint, which can be made stricter or looser by tuning a parameter. \cite{KaZh21} has an explicit decision problem (pricing), but similarly introduces a kind of budget constraint that allows for some tunable degree of unfairness. There is some difficulty in interpreting such solutions, as it is not clear how much unfairness should be considered ``acceptable.'' This is why, in our work, we first characterize an ideal fair policy, then approximate it in an optimal manner: the unfairness of the approximation vanishes to zero at a provably optimal rate.

A different approach is to transform the data $X$, or the performance $\beta^\top X$, in a way that removes the dependence on $Z$. Two pioneering papers on this topic are \cite{JoLu19} and \cite{Chzhen20}. Both adopt statistical parity as their fairness criterion and derive an ``ideal'' transformation that is provably fair when the problem inputs are known. In that sense, these studies are the closest to our work among all the literature surveyed here; furthermore, \cite{Chzhen20} proves that this transformation technique is optimal in the sense of minimizing prediction error. The crucial difference is that these papers do not formulate any decision problem. In Section \ref{sec:uisbad}, we give a counter-example proving that the ``cdf-transformation'' technique that gives optimal fair predictions is \textit{not} guaranteed to lead to optimal hiring decisions. The core methodological contribution of our paper is a fair \textit{hiring policy} that is provably optimal, not in the sense of predictive power, but with respect to the \textit{economic} objective of the employer. Our results show that these two types of problems have very different solutions.

\section{Fair decisions and their properties}

Let $\left(X,Z\right)$ be a sample from the data-generating process, with $X$ taking values in $\mathcal{X}\subseteq\mathds{R}^p$ and $Z$ taking values in $\left\{0,1\right\}$. Let $Y$ be related to $X$ through the linear regression model (\ref{eq:reg}), and let $\rho = P\left(Z=1\right)$ denote the mean of $Z$. Throughout this section, we treat $\beta$ and the joint distribution of $\left(X,Z\right)$ as being known. Our first goal is to identify a notion of fairness that is suitable for the hiring problem. Once we characterize fair decisions in this ideal setting, we can then (in Section \ref{sec:finitesample}) discuss how to estimate them in practice.


The hiring problem is formalized as follows. We receive $K$ independent job applications from the general population, that is, we observe i.i.d. samples $\left\{X^k,Z^k\right\}^K_{k=1}$ from the data-generating process. For notational convenience, let $\bm{X} = \left(X^1,...,X^K\right)$ and $\bm{Z} = \left(Z^1,...,Z^K\right)$. A hiring policy $\pi$ maps $\left(\bm{X},\bm{Z}\right)$ into $\left\{1,...,K\right\}$. To reduce notational clutter, we will often write the index $\pi\left(\bm{X},\bm{Z}\right)$ of the selected candidate as simply $\pi$.

We define the \textit{ideal fair policy} as
\begin{equation}\label{eq:ideal}
\pi^* = \arg\max_{\pi\in\Pi} \mathbb{E}\left(\beta^\top X^{\pi}\right),
\end{equation}
where $\Pi$ is the set of all measurable functions $\pi$ satisfying $\pi\left(\bm{X},\bm{Z}\right) \perp \bm{Z}$, with the symbol $\perp$ denoting independence. The independence condition can be equivalently written as $P\left(\pi=k\mid \bm{Z}\right) = P\left(\pi=k\right)$ for any $k=1,...,K$ and $\pi\in\Pi$. To rule out trivial policies such as $\pi\equiv 1$, we further require $P\left(\pi=k\right) \equiv\frac{1}{K}$, because the candidates are i.i.d. and, if nothing is known about them, any one should be equally likely to be chosen.\footnote[1]{Another way to interpret this condition is that the policy should be invariant with respect to permutations of the indices $1,...,K$.} The fairness condition says that this must still hold even if $\bm{Z}$ (but not $\bm{X}$) is known. At the same time, among all policies that are fair in this sense, we wish to select the one that maximizes expected performance. There may be many fair but suboptimal policies; Section \ref{sec:uisbad} presents a non-trivial example.

In the following, Section \ref{sec:noparity} explains why unconstrained maximization of $\mathbb{E}\left(\beta^\top X^{\pi}\right)$ causes serious ethical concerns in hiring, thus justifying the choice of (\ref{eq:ideal}) as the target. Section \ref{sec:uisbad} shows a fundamental difference between (\ref{eq:ideal}) and the well-studied fair prediction problem; specifically, we prove that it is not optimal to rank applicants by optimal fair predictions of their performance. Section \ref{sec:idealprop} then gives a complete characterization of $\pi^*$.

\subsection{Why we do not use parity of treatment}\label{sec:noparity}

It may not be obvious why we see the policy $\pi^{\max} = \arg\max_k \beta^\top X^k$ as being unfair. It selects the best-performing candidate and does not functionally depend on $\bm{Z}$, thus appearing to be blind to the protected attribute.\footnote[2]{For example, anonymizing r\'{e}sum\'{e}s so that hiring managers cannot see the names of applicants may be seen as an attempt to implement this policy in practice.} It also meets the parity of treatment criterion: $\pi^{\max}$ is conditionally independent of $\bm{Z}$ given $\bm{X}$.

What makes $\pi^{\max}$ unfair is the competitiveness of the job opening. Only one applicant will be hired. As $K$ increases, the outcome is decided by competition, \textit{not} between ``typical'' applicants in each subgroup, but between extreme values. This insight is made very stark by the following formal result, whose proof is deferred to the Appendix.

\begin{prop}\label{prop:extreme}
Let $F$ denote the (unconditional) cdf of $\beta^\top X$. Define
\begin{equation*}
\eta\left(q\right) = P\left(Z = 1 \mid F\left(\beta^\top X\right) = q\right).
\end{equation*}
Then,
\begin{eqnarray*}
\liminf_{K\rightarrow\infty} P\left(Z^{\pi^{\max}} = 1\right) &\geq & \liminf_{q\nearrow 1}\eta\left(q\right) = \lim_{q\nearrow 1}\inf_{q'\geq q}\eta\left(q'\right),\\
\limsup_{K\rightarrow\infty} P\left(Z^{\pi^{\max}} = 1\right) &\leq & \limsup_{q\nearrow 1}\eta\left(q\right) = \lim_{q\nearrow 1}\sup_{q'\geq q}\eta\left(q'\right).
\end{eqnarray*}
\end{prop}

To illustrate the import of Proposition \ref{prop:extreme}, consider the following example. Suppose, for the sake of argument, that the conditional distribution of $\beta^\top X$, given $Z = z$, is $\mathcal{N}\left(0,\tau^2_z\right)$ with $\tau_0 > \tau_1$ (the difference may be arbitrarily small). The average performance of a single individual is the same in either subgroup. However, Proposition \ref{prop:extreme} implies that $P\left(Z^{\pi^{\max}} = 1\right)\rightarrow 0$ as $K\rightarrow\infty$. Although more and more minority candidates are applying, the chance that any of them will be hired is vanishing.

The reliance of $\pi^{\max}$ on the tails of the data-generating distribution has the effect of amplifying disparities between the two groups. Even when the disparity appears to be small on an individual level (e.g., if $\tau_0$ is only slightly greater than $\tau_1$), it effectively prevents the disadvantaged group from being represented \textit{at all} when the applicant pool is large. We see this as a serious ethical concern with using parity of treatment in hiring.

\subsection{Why we do not use fair prediction}\label{sec:uisbad}

In the literature on fair \textit{prediction}, we have a single $\left(X,Z\right)$ pair (i.e., $K=1$) and wish to solve
\begin{equation}\label{eq:besttotalprediction}
\psi^* = \arg\min_{\psi\in\Psi}\mathbb{E}\left[\left(\beta^\top X - \psi\left(X,Z\right)\right)^2\right],
\end{equation}
where $\Psi$ is the set of all measurable functions $\psi$ satisfying $\psi\left(X,Z\right)\perp Z$. While this formulation has some resemblance to (\ref{eq:ideal}), the function $\psi$ is an \textit{estimator} of $\beta^\top X$, not a \textit{policy} that chooses between candidates; the goal is simply to estimate the performance of a single candidate in a fair way without any decision problem.

\cite{Chzhen20} showed that (\ref{eq:besttotalprediction}) has a closed-form solution. Denote by
\begin{equation*}
F_z\left(r\right) = P\left(\beta^\top X \leq r\,|\,Z = z\right), \qquad z\in\left\{0,1\right\}
\end{equation*}
the conditional cdf of the performance $\beta^\top X$ of an applicant with attributes $X$, given $Z = z$. Then, the random variable $U = F_Z\left(\beta^\top X\right)$ follows a uniform distribution on $\left[0,1\right]$ and is independent of $Z$. This quantity represents the percentile rank of the applicant relative to their subgroup (on the population level). The optimal fair prediction is given by
\begin{equation*}
\psi^*\left(U\right) = \rho F^{-1}_1\left(U\right) + \left(1-\rho\right)F^{-1}_0\left(U\right),
\end{equation*}
a simple weighted average of the two conditional inverse cdfs, and a monotonic function of $U$.

A natural extension of this idea to the hiring problem is the simple policy $\pi^U = \arg\max_k U^k$, where $U^k = F_{Z^k}\left(\beta^\top X^k\right)$, which ranks applicants based on optimal fair predictions of their performance. Clearly $\pi^U \perp \bm{Z}$ and $P\left(\pi^U=k\right)\equiv \frac{1}{K}$, so the fairness condition is satisfied. It turns out, however, that $\pi^U$ is not optimal, as shown in the following counter-example.

\begin{prop}
There exists an instance of the hiring problem where $\pi^* \neq \pi^U$.
\end{prop}

\noindent\textbf{Proof:} Take $K=2$, and let $F^{-1}_0\left(u\right) = \frac{3}{8}+\frac{1}{4}u$ and $F^{-1}_1\left(u\right) = u$. Suppose that $\bm{Z}=\left(0,1\right)$ is given. Then,
\begin{eqnarray*}
\mathbb{E}\left(\beta^\top X^{\pi^U}\mid \bm{Z}=\left(0,1\right)\right) &=& \mathbb{E}\left(F^{-1}_0\left(U^1\right)1_{\left\{U^1\geq U^2\right\}} + F^{-1}_1\left(U^2\right)1_{\left\{U^1< U^2\right\}}\mid \bm{Z}=\left(0,1\right)\right)\\
&=& \mathbb{E}\left(F^{-1}_1\left(U^2\right) + \left(F^{-1}_0\left(U^1\right)-F^{-1}_1\left(U^2\right)\right)1_{\left\{U^1\geq U^2\right\}}\right)\\
&=& \frac{1}{2} + \mathbb{E}\left(\left(\frac{3}{8}+\frac{1}{4}U^1\right)1_{\left\{U^1\geq U^2\right\}}\right) - \mathbb{E}\left(U^2 1_{\left\{U^1\geq U^2\right\}}\right).
\end{eqnarray*}
We then calculate
\begin{eqnarray*}
\mathbb{E}\left(\left(\frac{3}{8}+\frac{1}{4}U^1\right)1_{\left\{U^1\geq U^2\right\}}\right) &=& \mathbb{E}\left(\mathbb{E}\left(\left(\frac{3}{8}+\frac{1}{4}U^1\right)1_{\left\{U^1\geq U^2\right\}}\mid U^1\right)\right)\\
&=& \mathbb{E}\left(\frac{3}{8}U^1 + \frac{1}{4}\left(U^1\right)^2\right)\\
&=& \frac{13}{48}.
\end{eqnarray*}
Analogously,
\begin{equation*}
\mathbb{E}\left(U^2 1_{\left\{U^1\geq U^2\right\}}\right) = \mathbb{E}\left(\mathbb{E}\left(U^2 1_{\left\{U^1\geq U^2\right\}}\right)\mid U^2\right)= \mathbb{E}\left(U^2 - \left(U^2\right)^2\right)= \frac{1}{6},
\end{equation*}
whence $\mathbb{E}\left(\beta^\top X^{\pi^U}\mid \bm{Z}=\left(0,1\right)\right) = \frac{29}{48}$.

Now define another policy $\pi$ as follows. When $\bm{Z} \neq \left(0,1\right)$, the policy is identical to $\pi^U$. When $\bm{Z} = \left(0,1\right)$, the policy chooses candidate $2$ if $U^2 > \frac{1}{2}$, and candidate $1$ otherwise. This policy satisfies the fairness condition because $P\left(\pi=1\mid\bm{Z}\right)=\frac{1}{2}$ for all $\bm{Z}$. Then, similarly to the previous computations,
\begin{equation*}
\mathbb{E}\left(\beta^\top X^{\pi}\mid \bm{Z}=\left(0,1\right)\right) = \frac{1}{2} + \mathbb{E}\left(\left(\frac{3}{8}+\frac{1}{4}U^1\right)1_{\left\{U^2<\frac{1}{2}\right\}}\right) - \mathbb{E}\left(U^21_{\left\{U^2< \frac{1}{2}\right\}}\right)
\end{equation*}
where
\begin{equation*}
\mathbb{E}\left(\left(\frac{3}{8}+\frac{1}{4}U^1\right)1_{\left\{U^2<\frac{1}{2}\right\}}\right) = \frac{1}{2}\mathbb{E}\left(\frac{3}{8}+\frac{1}{4}U^1\right) = \frac{1}{4}
\end{equation*}
and $\mathbb{E}\left(U^21_{\left\{U^2\leq \frac{1}{2}\right\}}\right) = \frac{1}{8}$. Thus, $\mathbb{E}\left(\beta^\top X^{\pi}\mid \bm{Z}=\left(0,1\right)\right) = \frac{5}{8}$, which is greater than the value obtained by $\pi^U$. Therefore, $\pi^U$ is not optimal.\qed

Ranking candidates by their percentile within subgroup is not optimal because (as is also the case in the counter-example) the conditional distributions $F_z$ may have very different tails, and a lower percentile in one subgroup may actually be farther out on the positive tail of performance. As we will see next, the \textit{optimal} fair policy directly compares the best-performing candidates from each subgroup.

\subsection{Characterization of the ideal fair policy}\label{sec:idealprop}

For convenience, let us introduce the notation
\begin{eqnarray*}
K_z &=& \sum^K_{k=1} 1_{\left\{Z^k=z\right\}}, \qquad z\in\left\{0,1\right\},\\
R^{(z)} &=& \max_{k:Z^k = z} \beta^\top X^k, \qquad z\in\left\{0,1\right\}.
\end{eqnarray*}
These quantities can be computed from $\left(\bm{X},\bm{Z}\right)$ and thus are known to the policy when the decision is made. Without loss of generality, suppose that the policy is given a set of candidates with $K_0,K_1 > 0$ (if all the candidates belong to the same subgroup, the decision is trivial).

For any fair policy $\pi$, we can write
\begin{eqnarray}
\beta^\top X^\pi &=& \sum_k \beta^\top X^k 1_{\left\{\pi = k\right\}}\nonumber\\
&\leq & R^{(1)} 1_{\left\{Z^{\pi}=1\right\}} + R^{(0)} 1_{\left\{Z^{\pi}=0\right\}}\nonumber\\
&=& R^{(0)} + 1_{\left\{Z^{\pi}=1\right\}}\cdot \left(R^{(1)}-R^{(0)}\right).\label{eq:vpi}
\end{eqnarray}
Intuitively, this bound suggests that problem (\ref{eq:ideal}) can be simplified by maximizing (\ref{eq:vpi}) directly. The idea (which will be justified rigorously further down) is that it is sufficient to compare two candidates, namely $\arg\max_{k:Z^k=1} \beta^\top X^k$ and $\arg\max_{k:Z^k=0} \beta^\top X^k$, and the only question is when to choose which subgroup.

In the absence of any constraints, we would follow the parity of treatment policy, which indeed has this structure, with $\left\{Z^{\pi^{\max}}=1\right\}=\left\{R^{(1)}\geq R^{(0)}\right\}$. The fairness condition, however, requires
\begin{equation}\label{eq:proportion}
P\left(Z^{\pi} = z\mid \bm{Z}\right) = \frac{K_z}{K}, \qquad z\in\left\{0,1\right\},
\end{equation}
which straightforwardly follows from $P\left(\pi=k\mid\bm{Z}\right)\equiv \frac{1}{K}$. Intuitively, if we are comparing the top candidates from each subgroup, we should choose group $1$ on a set of outcomes covering a proportion $\frac{K_1}{K}$ of the largest values of $R^{(1)}-R^{(0)}$. The following result confirms our intuition.

\begin{thm}\label{thm:ideal}
The fair policy $\pi^*$ that maximizes (\ref{eq:ideal}) is given by
\begin{equation}\label{eq:idealclosedform}
\pi^* = \left\{
\begin{array}{c l}
\arg\max_{k:Z^k=1} \beta^\top X^k & R^{(1)}- R^{(0)}\geq q\left(\bm{Z}\right),\\
\arg\max_{k:Z^k=0} \beta^\top X^k & \mbox{otherwise,}
\end{array}
\right.
\end{equation}
where $q\left(\bm{Z}\right)$ satisfies
\begin{equation*}
P\left(R^{(1)}- R^{(0)}\geq q\left(\bm{Z}\right)\mid \bm{Z}\right) = \frac{K_1}{K}.
\end{equation*}
In words, $q\left(\bm{Z}\right)$ is the $\frac{K_0}{K}$-quantile of the \textit{conditional} distribution of $R^{(1)}- R^{(0)}$ given $\bm{Z}$.
\end{thm}

\noindent\textbf{Proof:} Pick any fair policy $\pi$ and let $A = \left\{Z^{\pi}=1\right\}$. By (\ref{eq:proportion}), we have $P\left(A\mid\bm{Z}\right) = \frac{K_1}{K}$. We compare $A$ with the set
\begin{equation*}
A^* = \left\{R^{(1)}- R^{(0)}\geq q\left(\bm{Z}\right)\right\},
\end{equation*}
which satisfies $P\left(A^*\mid \bm{Z}\right)=\frac{K_1}{K}$ by construction. Note that $\pi^*$ satisfies the fairness condition $P\left(\pi^*=k\mid\bm{Z}\right)\equiv\frac{1}{K}$ because, conditionally on $\bm{Z}$, the policy selects subgroup $z\in\left\{0,1\right\}$ w.p. $\frac{K_z}{K}$, and each individual within that subgroup has the same conditional probability $\frac{1}{K_z}$ of being the best, independently of $R^{(z)}$.

For policy $\pi^*$, the bound (\ref{eq:vpi}) holds with equality, and
\begin{eqnarray*}
\beta^\top\left(X^{\pi^*}-X^{\pi}\right) &\geq& \left(R^{(1)}-R^{(0)}\right)\cdot\left(1_{A^*} - 1_{A}\right)\\
&=& \left(R^{(1)}-R^{(0)}\right)\cdot\left(1_{A^*\cap A}+1_{A^*\cap A^c} - 1_{A\cap A^*}-1_{A\cap \left(A^*\right)^c}\right)\\
&=& \left(R^{(1)}-R^{(0)}\right)\cdot\left(1_{A^*\cap A^c} - 1_{A\cap \left(A^*\right)^c}\right).
\end{eqnarray*}
Taking conditional expectations, we obtain
\begin{eqnarray*}
\mathbb{E}\left(\left(R^{(1)}-R^{(0)}\right) 1_{A\cap \left(A^*\right)^c}\mid\bm{Z}\right) &\leq& q\left(\bm{Z}\right)\cdot P\left(A\cap \left(A^*\right)^c\mid\bm{Z}\right)\\
&=& q\left(\bm{Z}\right)\cdot P\left(A^*\cap A^c\mid\bm{Z}\right)\\
&\leq& \mathbb{E}\left(\left(R^{(1)}-R^{(0)}\right) 1_{A^*\cap A^c}\mid\bm{Z}\right),
\end{eqnarray*}
where the second line follows because $P\left(A\mid\bm{Z}\right)=P\left(A^*\mid\bm{Z}\right)=\frac{K_1}{K}$. Then, $\mathbb{E}\left(\beta^\top\left(X^{\pi^*}-X^{\pi}\right)\mid \bm{Z}\right) \geq 0$, whence $\mathbb{E}\left(\beta^\top\left(X^{\pi^*}-X^{\pi}\right)\right) \geq 0$, as desired.\qed

From (\ref{eq:idealclosedform}), we see that the ideal fair policy directly compares the best-performing candidates from each subgroup. The decision is made based on the difference $R^{(1)}-R^{(0)}$, which favors minority candidates both when they are performing well and when non-minority candidates are under-performing. This difference must be above the threshold $q\left(\bm{Z}\right)$, which ensures that we do not hire under-qualified candidates only to satisfy the fairness constraint. However, to correct for the effects of systemic and structural bias, this threshold is adjusted, and may be negative in some problem instances. Note also that (\ref{eq:idealclosedform}) is completely different from the optimal fair prediction discussed in Section \ref{sec:uisbad}, which starkly illustrates the distinction between fair prediction and fair decisions.

Theorem \ref{thm:ideal} gives us the optimal policy in an ideal situation where the underlying model is known, and thus gives us a standard to aim for. Computing this policy can be quite difficult, as we would then need the $\frac{K_0}{K}$-quantile of a fairly complicated distribution for each possible value of $\bm{Z}$. However, when $\beta$ and $F_z$ are no longer known, it is only necessary to \textit{estimate} the policy, which turns out to be a much more tractable task.

\section{Finite-sample policy and analysis}\label{sec:finitesample}

We will now suppose that $\beta$ and $F_z$ are unknown, but can be estimated from historical data. These data are not the same as the information $\left(\bm{X},\bm{Z}\right)$ about the pool of $K$ candidates from which the selection decision $\pi$ is made. We now suppose that we have access to
\begin{equation*}
\bm{X}_n = \left[
\begin{array}{c c c}
X_{1,1} & \ldots & X_{1,p}\\
\vdots & \ddots & \vdots\\
X_{n,1} & \ldots & X_{n,p}
\end{array}
\right], \qquad \bm{Z}_n = \left[
\begin{array}{c}
Z_1\\
\vdots\\
Z_n
\end{array}
\right], \qquad \bm{Y}_n = \left[
\begin{array}{c}
Y_1\\
\vdots\\
Y_n
\end{array}
\right].
\end{equation*}
For each $m \leq n$, the pair $\left(X_m,Z_m\right)$ with $X_m = \left(X_{m,1},...,X_{m,p}\right)$ is an i.i.d. sample from the data-generating distribution, with $Y_m$ obtained from (\ref{eq:reg}) applied to $X_m$ and an i.i.d. noise $\varepsilon_m$. The response variable is only observed for the historical data points, and not for the data $\left\{X^k,Z^k\right\}^K_{k=1}$ describing the new candidates.

Note that $\bm{X}_n$ and $\bm{Z}_n$, which were observed in the past, are distinct (and independent) from $\bm{X}$ and $\bm{Z}$, which have just now been observed. In order to visually distinguish between the historical and new data, we use $k$ in the superscript to denote one of the new data points, and $m$ or $n$ in the subscript to denote a historical one. Thus, $X^k$ is a $p$-vector containing the features of the $k$th new candidate, while $X_m$ is the $m$th row of the historical data matrix $\bm{X}_n$. For shorthand, when it is necessary to refer to all of the historical data together, we write $\bm{H}_n = \left(\bm{X}_n,\bm{Z}_n,\bm{Y}_n\right)$. The definition of a policy is now extended to functions $\pi$ that map $\left(\bm{H}_n,\bm{X},\bm{Z}\right)$ into $\left\{1,...,K\right\}$, so that the decision is now based on the historical data.

In this paper, we assume that the decision is only made once, i.e., the problem is clearly separated into a training stage (using the historical data) followed by a decision stage (given $\bm{X}$ and $\bm{Z}$). Such a setup is sometimes called ``offline.'' Consequently, in our notation we will suppress the dependence of policies on the sample size $n$, since the dataset is fixed. This two-stage model is a natural first step for developing a fundamental understanding of fairness in decision-making; other recent work on prescriptive analytics, such as \cite{BeKa20}, also separates training and decision-making in a similar way. We do not consider online (sequential) decision-making in this work, but the results of \cite{BaBaKh21} suggest that this setting can be handled relatively easily as long as the data-generating process is sufficiently rich.

Section \ref{sec:impossible} first fixes a finite $n$ and proves that any non-trivial $\pi$ constructed from the training data cannot be guaranteed to satisfy $\pi\left(\bm{H}_n,\bm{X},\bm{Z}\right) \perp \bm{Z}$. In other words, it is not possible to be completely fair in finite sample. This motivates the study of an empirical approximation of (\ref{eq:idealclosedform}), formally stated in Section \ref{sec:empirical}. This empirical policy is also not completely fair, but we prove that it is the best possible approximation of the ideal fair policy: Section \ref{sec:empirical} gives an upper bound on the asymptotic convergence rate of the probability that the empirical policy deviates from (\ref{eq:idealclosedform}), and Section \ref{sec:minimax} proves a lower bound of the same order.

\subsection{Fairness is unachievable in finite sample}\label{sec:impossible}

The main result of this subsection is that, in finite sample, any perfectly fair policy $\pi\left(\bm{H}_n,\bm{X},\bm{Z}\right)$ cannot depend on $\bm{X}$. In other words, the only possible fair policies in finite sample are trivial ones, e.g., the policy that always selects the index of the first candidate.

Without loss of generality, suppose $K=2$, so $\bm{X} = \left(X^1,X^2\right)$ and $\bm{Z} = \left(Z^1,Z^2\right)$. Let $\pi$ be a policy that is fair in finite sample, i.e., satisfies $\pi\left(\bm{H}_n,\bm{X},\bm{Z}\right) \perp \bm{Z}$. For any $z =\left(z^1,z^2\right)$ with $z^1,z^2\in\left\{0,1\right\}$, define
\begin{equation*}
A\left(z^1,z^2,\bm{H}_n\right) = \left\{ x\,:\, \pi\left(\bm{H}_n,x,z\right) = 1\right\}
\end{equation*}
to be the set of all possibly new data points that would lead the policy to select the first candidate under the given, fixed $z$ values. The fairness condition on $\pi$ requires
\begin{equation}\label{eq:aproperty}
P\left( \bm{X} \in A\left(1,1,\bm{H}_n\right)\mid \bm{Z}=\left(1,1\right),\bm{H}_n\right) = P\left( \bm{X} \in A\left(0,1,\bm{H}_n\right)\mid \bm{Z}=\left(0,1\right),\bm{H}_n\right)
\end{equation}
under \textit{any} choice of conditional distribution of $\bm{X}$ given $\bm{Z}$. Note that, conditional on $\bm{H}_n$, both $A\left(1,1,\bm{H}_n\right)$ and $A\left(0,1,\bm{H}_n\right)$ are known (can be treated as fixed sets).

Let $g_z$ denote the conditional density of $X$ given $Z = z$ for $z\in\left\{0,1\right\}$. Then, recalling that $\left(X^1,Z^1\right) \perp \left(X^2,Z^2\right)$, (\ref{eq:aproperty}) is equivalent to the requirement
\begin{equation}\label{eq:adoubleint}
\iint 1_{\left\{\left(x^1,x^2\right)\in A\left(1,1,\bm{H}_n\right)\right\}}g_1\left(x^1\right)g_1\left(x^2\right)dx^1 dx^2 = \iint 1_{\left\{\left(x^1,x^2\right)\in A\left(0,1,\bm{H}_n\right)\right\}}g_0\left(x^1\right)g_1\left(x^2\right)dx^1 dx^2
\end{equation}
for any choice of $g_0,g_1$. The key to the result is that, given fixed data $\bm{H}_n$, the sets $A\left(1,1,\bm{H}_n\right)$ and $A\left(0,1,\bm{H}_n\right)$ in (\ref{eq:adoubleint}) cannot have any additional dependence on the densities $g_0,g_1$. Then, the only way to make the fairness condition (\ref{eq:adoubleint}) hold for all possible data-generating distributions is to have $\pi$ ignore $\bm{X}$ completely. If we knew the data-generating distribution exactly, then we could design $\pi$ to explicitly depend on $g_0,g_1$, but this is not allowed in finite sample.

To state the result more precisely, if $Z^2 = 1$ and $\pi$ picks the first candidate with nonzero probability, then it must always pick the first candidate (regardless of $\bm{X}$). We can also obtain the same result with $Z^2 = 0$ by rewriting (\ref{eq:adoubleint}) with $A\left(1,0,\bm{H}_n\right)$ and $A\left(0,0,\bm{H}_n\right)$, and repeating the same arguments.

\begin{thm}\label{thm:xandj}
Suppose that (\ref{eq:adoubleint}) holds for all $g_0,g_1$, and suppose that $P\left(\bm{X}\in A\left(0,1,\bm{H}_n\right)\right) > 0$. Then, $A\left(0,1,\bm{H}_n\right) = A\left(1,1,\bm{H}_n\right) = \mathcal{X}\times\mathcal{X}$.
\end{thm}

The implication of Theorem \ref{thm:xandj} is that, for any finite sample size $n$, it is not possible to achieve near-optimal (or even good) empirical performance and simultaneously maintain perfect fairness. This motivates the study of empirical policies that are not perfectly fair in finite sample, but are guided by the ideal fair policy, i.e., approximate it as closely as possible given the data. In the remainder of this section, we present one such policy $\hat{\pi}$ and show that $P\left(\hat{\pi}\neq\pi^*\right)$ converges to zero at the fastest possible rate.

\subsection{Empirical policy and performance bound}\label{sec:empirical}

We propose to approximate (\ref{eq:idealclosedform}) by
\begin{equation}\label{eq:pihat}
\hat{\pi} = \left\{
\begin{array}{c l}
\arg\max_{k:Z^k=1} \hat{\beta}^\top X^k & \hat{R}^{(1)}- \hat{R}^{(0)}\geq \hat{q}\left(\bm{Z}\right),\\
\arg\max_{k:Z^k=0} \hat{\beta}^\top X^k & \mbox{otherwise.}\\
\end{array}
\right.
\end{equation}
where
\begin{equation*}
\hat{R}^{(z)} = \max_{k:Z^k = z} \hat{\beta}^\top X^k, \qquad z\in\left\{0,1\right\},
\end{equation*}
and $\hat{\beta},\hat{q}\left(\bm{Z}\right)$ are estimators of $\beta,q\left(\bm{Z}\right)$ constructed from the data $\bm{H}_n$. For the first quantity, we use the ordinary least squares estimator
\begin{equation*}
\hat{\beta} = \left(\bm{X}^\top_n \bm{X}_n\right)^{-1} \bm{X}^\top_n \bm{Y}_n.
\end{equation*}
To construct the estimated quantile $\hat{q}\left(\bm{Z}\right)$, first recall that $q\left(\bm{Z}\right)$ is the $\frac{K_0}{K}$-quantile of the conditional distribution of $R^{(1)}-R^{(0)}$ given $\bm{Z}$. Note that we know $\bm{Z}$, and thus $K_0,K_1$, when the hiring decision is made. Then, we can let $\hat{q}\left(\bm{Z}\right)$ be the quantile of the corresponding \textit{empirical} distribution. That is, for $z \in\left\{0,1\right\}$, let $m_1,...,m_{K_z}$ be $K_z$ indices, drawn uniformly with replacement, from the set $\left\{m\leq n:Z_m=z\right\}$, and let $\tilde{R}^{(z)} = \max_{k=1,...,K_z} \hat{\beta}^\top X_{m_k}$. The data (and therefore the estimator $\hat{\beta}$) are fixed, so if $\bm{Z}$ is given, these indices are the only source of randomness in $\tilde{R}^{(z)}$. Then, $\hat{q}\left(\bm{Z}\right)$ is the $\frac{K_0}{K}$-quantile of the distribution of $\tilde{R}^{(1)}-\tilde{R}^{(0)}$.

This empirical quantile can be estimated very easily using the bootstrap method \citep{EfTi94}, as one can repeatedly sample the indices $m_1,...,m_{K_z}$ and calculate samples of $\tilde{R}^{(1)}-\tilde{R}^{(0)}$, then take the $\frac{K_0}{K}$-quantile of these bootstrapped values. This is sufficient for practical implementation. However, it is also possible to compute $\hat{q}\left(\bm{Z}\right)$ exactly, and we give an algorithm for doing so in the Appendix. Our theoretical analysis uses the exact value of $\hat{q}\left(\bm{Z}\right)$.

The empirical policy $\hat{\pi}$ is not perfectly fair, but from Section \ref{sec:impossible} we also know that it is impossible for a nontrivial policy to be fair in finite sample. We evaluate the relative fairness of $\hat{\pi}$ in terms of the probability $P\left(\hat{\pi} \neq \pi^*\right)$ that $\hat{\pi}$ deviates from (\ref{eq:idealclosedform}). We first prove an upper bound on this probability, and provide a lower bound of the same order in Section \ref{sec:minimax}.

The assumptions on the data-generating process required for the bound are stated below. By and large, they are quite standard; for example, we require the matrix $\mathbb{E}\left(XX^\top\right)$ to be positive definite to ensure consistency of the OLS estimator $\hat{\beta}$. The most important assumptions are that the residual noise $\varepsilon$ is sub-Gaussian and the data $X$ are bounded. Both of these conditions are quite common in recent work on data-driven decision-making in the management science community; examples include, but are not limited to, \cite{BaBa20} and \cite{BaBaKh21}.

\begin{assum}\label{a1}
There exist finite values $\kappa_i > 0$, $i=1,...,8$, such that the following conditions are satisfied:
\begin{itemize}
\item[i)] $\mathbb{E}\left(Z\right) \in \left[\kappa_1,1-\kappa_1\right]$.
\item[ii)] All eigenvalues of the matrix $\mathbb{E}\left(XX^\top\right)$ are in the interval $\left[\kappa_2,\kappa_3\right]$.
\item[iii)] The data $X$ are bounded, i.e., $P\left(\|X\|_2 \leq \kappa_4\right) = 1$.
\item[iv)] The sub-Gaussian norm of the residual noise $\varepsilon$, given by
\begin{equation*}
\|\varepsilon\|_{SG} =\inf\left\{t>0\,:\,\mathbb{E}\left(\exp\left(\frac{\varepsilon^2}{t^2}\right)\right)\leq 2\right\},
\end{equation*}
satisfies $\|\varepsilon\|_{SG} \leq \kappa_5$.
\item[v)] For any $z\in\left\{0,1\right\}$, the conditional distribution of $X$ given $Z=z$ has a density that is bounded by $\kappa_6$.
\item[vi)] The conditional density of $R^{(1)}-R^{(0)}$ given $\bm{Z}$ is bounded below by $\kappa_8$ on the interval $q\left(\bm{Z}\right)\pm \kappa_7$.
\end{itemize}
\end{assum}

There are two versions of the upper bound: the first is conditional on $\left(\bm{X},\bm{Z}\right)$ and has an exponential rate, and the second is unconditional (averaged over all possible new data) and has the rate $\mathcal{O}\left(n^{-\frac{1}{2}}\right)$. For readability, we will state and discuss the bounds here. Section \ref{sec:upperbound} provides further details on the key arguments of the proofs.

\begin{thm}\label{thm:upperbound}
Under Assumption \ref{a1}, there exist problem-specific constants $C,C',C''$ such that
\begin{eqnarray}
P\left(\hat{\pi} \neq \pi^*\mid\bm{X},\bm{Z}\right) &\leq& C \exp\left(-C' n \cdot g\left(\bm{X},\bm{Z}\right)\right),\label{eq:conditional}\\
P\left(\hat{\pi} \neq \pi^*\right) &\leq& C'' n^{-\frac{1}{2}}.\label{eq:unconditional}
\end{eqnarray}
where $g\left(\bm{X},\bm{Z}\right) > 0$ depends on $\left(\bm{X},\bm{Z}\right)$, but not on $n$.
\end{thm}

Thus, given any \textit{fixed} group of candidates, the empirical policy converges to the ideal fair policy exponentially, but the rate exponent is a function of $\left(\bm{X},\bm{Z}\right)$. In the unconditional version, this exponent varies randomly, and may be arbitrarily small, making it more difficult to rank the candidates on average over all possible realizations from the data-generating process.

The distinction between the conditional and unconditional bounds is noteworthy because it does not arise in other problem classes where decisions are ranked using estimated values, such as ordinal optimization \citep{ShBrZe18}. In many such problems, the \textit{true} values of the decisions are fixed (though unknown to the decision-maker), and the only noise comes from estimation error. In such problems, one can often obtain exponential decay rates for error probabilities, but the precise exponents also depend on the values \citep{Ru20}. In the present setting, these values are not fixed, but are determined by sampling from the data-generating process, which introduces additional noise. Essentially, the policy now has to separate the estimation error from the variation in the candidate pool, so it is not surprising that the convergence rate becomes slower.

\subsection{Minimax lower bound}\label{sec:minimax}

We now establish a lower (minimax) bound of $\mathcal{O}\left(n^{-\frac{1}{2}}\right)$ on the probability of deviating from the fair decision. The precise meaning of such a bound is that, for any empirical policy, one can always find at least one data-generating process that makes the deviation probability vanish at the indicated rate. We may impose additional restrictions on the data-generating process without loss of generality -- if we can find a sufficiently ``bad'' distribution from the restricted class, the ensuing lower bound will still hold for a more general class. To put it another way, the worst case can only become worse if a more general class of data-generating processes is permitted.

Since we are free to impose additional structure on the data-generating process, let us assume that $X \perp Z$ with $\mathbb{E}\left(Z\right) = \frac{1}{2}$ and $X\sim \mathcal{N}\left(0,\bm{I}_p\right)$, where $\bm{I}_p$ is the $p\times p$ identity matrix. Suppose that $Y$ is related to $X$ through (\ref{eq:reg}) with $\varepsilon\sim\mathcal{N}\left(0,\sigma^2\right)$. Let us fix $\sigma > 0$ to an arbitrary strictly positive value. Thus, the distribution of $\left(\bm{H}_n,\bm{X},\bm{Z}\right)$ is parameterized by $\beta$. In our proof we will further restrict ourselves to $\beta \in \left\{\beta_0,\beta_1\right\}$ for suitable choices of $\beta_0,\beta_1$.

Without loss of generality, suppose that $K=2$, so $\bm{X} = \left(X^1,X^2\right)$ and $\bm{Z} = \left(Z^1,Z^2\right)$. Because $F_0=F_1$ in this setting, the ideal fair policy (\ref{eq:idealclosedform}) reduces to the indicator function $I_{\beta}\left(\bm{X}\right) = 1_{\left\{\beta^\top X^1 \geq \beta^\top X^2\right\}}$. We can now state the minimax bound (the proof is given in the Appendix).

\begin{thm}\label{thm:minimax}
Let $\Phi$ be the set of all functions mapping $\left(\bm{H}_n,\bm{X},\bm{Z}\right)$ into $\left\{0,1\right\}$. Then, there exist $\beta_0,\beta_1$ such that
\begin{equation*}
\inf_{\phi\in\Phi} \max_{\beta\in\left\{\beta_0,\beta_1\right\}} P_{\beta}\left(\phi\left(\bm{H}_n,\bm{X},\bm{Z}\right)\neq I_\beta\left(\bm{X}\right)\right) \gtrsim n^{-\frac{1}{2}},
\end{equation*}
where $P_{\beta}$ denotes a probability calculated using $\beta$ as the coefficients in model (\ref{eq:reg}).
\end{thm}

Theorem \ref{thm:minimax} shows that it is not possible to improve on the $\mathcal{O}\left(n^{-\frac{1}{2}}\right)$ rate obtained in (\ref{eq:unconditional}) for the empirical policy $\hat{\pi}$. In other words, $\hat{\pi}$ converges to the ideal fair policy $\pi^*$ at the best possible rate with respect to the sample size $n$. Together with the relative ease of computing $\hat{\pi}$, this constitutes a strong argument for the practical implementation of the empirical policy.

\section{Proof of upper bound}\label{sec:upperbound}

In this section, we give the main arguments justifying Theorem \ref{thm:upperbound}. Assumption \ref{a1} is assumed throughout. We use the notation $F_{z,\theta}\left(r\right) = P\left(\theta^\top X \leq r\,|\,Z=z\right)$ to distinguish between conditional cdfs obtained under different values $\theta$ for the regression coefficients $\beta$.

Let $\bm{Z}$ be given and suppose that $K_0,K_1 > 1$, as the policy becomes trivial otherwise. For $z\in\left\{0,1\right\}$, the conditional cdf of $R^{(z)}$ given $\bm{Z}$ is denoted by $G_{z,\beta} = \left(F_{z,\beta}\right)^{K_z}$. Then, for any $t$,
\begin{eqnarray*}
P\left(R^{(1)}-R^{(0)} \leq t\mid \bm{Z}\right) &=& \int G_{1,\beta}\left(u+t\right)dG_{0,\beta}\left(u\right)\\
&=& K_0 \int \left(F_{1,\beta}\left(u+t\right)\right)^{K_1}\left(F_{0,\beta}\left(u\right)\right)^{K_0-1} dF_{0,\beta}\left(u\right).
\end{eqnarray*}
Viewing $K_0,K_1$ as fixed and letting
\begin{equation}\label{eq:T}
T\left(t\right) = \int \left(F_{1,\beta}\left(u+t\right)\right)^{K_1}\left(F_{0,\beta}\left(u\right)\right)^{K_0-1} dF_{0,\beta}\left(u\right) - \frac{1}{K},
\end{equation}
we see that the desired quantile is the value $q$ satisfying $T\left(q\right) = 0$.

Let $n_z = \sum^n_{m=1} 1_{\left\{Z_m=z\right\}}$ and replace $F_{z,\beta}$ by the \textit{empirical} cdf
\begin{equation*}
\hat{F}_z\left(r\right) = \frac{1}{n_z}\sum^n_{m=1} 1_{\left\{\hat{\beta}^\top X_m \leq r\right\}}1_{\left\{Z_m=z\right\}}.
\end{equation*}
Define
\begin{equation}\label{eq:That}
\hat{T}\left(t\right) = \frac{1}{n_0}\sum^n_{m=1} \left(\hat{F}_1\left(\hat{\beta}^\top X_m + t\right)\right)^{K_1}\left(\hat{F}_0\left(\hat{\beta}^\top X_m\right)^{K_0-1}\right) 1_{\left\{Z_m=0\right\}} - \frac{1}{K}.
\end{equation}
We can take $\hat{q} = \arg\min_t \left|\hat{T}\left(t\right)\right|$. Given enough data, this quantity will converge to $q$.

From (\ref{eq:pihat}), we see that there are two basic ways in which the empirical policy $\hat{\pi}$ may fail to match the ideal fair policy $\pi^*$. First, $\hat{\pi}$ may fail to select the same subgroup as $\pi^*$. Second, $\hat{\pi}$ may select the correct subgroup, but choose the wrong candidate from this group due to estimation error in $\hat{\beta}$. Therefore,
\begin{eqnarray}
P\left(\hat{\pi}\neq\pi^*\mid \bm{X},\bm{Z}\right) &\leq& \sum_{z\in\left\{0,1\right\}} P\left(\arg\max_{k:Z^k=z} \hat{\beta}^\top X^k \neq \arg\max_{k:Z^k=z}\beta^\top X^k\mid \bm{X},\bm{Z}\right)\nonumber\\
&\,& + P\left(1_{\left\{\hat{R}^{(1)}- \hat{R}^{(0)}\geq \hat{q}\right\}}\neq 1_{\left\{R^{(1)}- R^{(0)}\geq q\right\}}\mid \bm{X},\bm{Z}\right).\label{eq:twotypes}
\end{eqnarray}
Then, to obtain the desired result, it is sufficient to bound each term on the right-hand side of (\ref{eq:twotypes}) by an exponentially decaying quantity of the form (\ref{eq:conditional}). The exponents do not need to match, since we can simply take $g\left(\bm{X},\bm{Z}\right)$ in (\ref{eq:conditional}) to be the smallest among them. The same is true for the unconditional bound (\ref{eq:unconditional}), as we may simply take expectations of both sides of (\ref{eq:twotypes}).

Conceptually, each of the two types of error occurs due to error in the relevant estimated parameters: within-subgroup error is caused by estimation error in $\hat{\beta}$, whereas subgroup selection error is caused by estimation error in both $\hat{\beta}$ and $\hat{q}$. The latter quantity, in turn, is derived from the estimators $\hat{F}_z$ and $\hat{T}$. Thus, we first have to establish sufficiently strong exponential decay rates on each type of estimation error. We then connect the two types of wrong decisions in (\ref{eq:twotypes}) to these concentration inequalities to arrive at the desired bounds.

\subsection{Bound on error within subgroup}\label{sec:estimationbound}

Take $z\in\left\{0,1\right\}$ and let $j_1,j_2$ be the indices of the largest and second-largest values in the set $\left\{\beta^\top X^k: Z^k = z\right\}$. We derive
\begin{eqnarray*}
&\,&\hspace{-0.1in} \left\{ \arg\max_{k:Z^k=z} \hat{\beta}^\top X^k \neq \arg\max_{k:Z^k=z}\beta^\top X^k\right\}\\
&=&\hspace{-0.1in} \left\{ \hat{\beta}^\top X^{j_1} \leq \max_{k:Z^k=z,k\neq j_1}\hat{\beta}^\top X^k\right\}\\
&=&\hspace{-0.1in} \left\{\beta^\top X^{j_1} + \left(\hat{\beta}-\beta\right)^\top X^{j_1}\leq \max_{k:Z^k=z,k\neq j_1}\beta^\top X^k + \left(\hat{\beta}-\beta\right)^\top X^k\right\}\\
&\subseteq &\hspace{-0.1in} \left\{ \beta^\top X^{j_1} - \max_{k'}\left|\left(\hat{\beta}-\beta\right)^\top X^k\right|\leq \beta^\top X^{j_2} + \max_{k'}\left|\left(\hat{\beta}-\beta\right)^\top X^k\right|\right\}\\
&\subseteq &\hspace{-0.1in} \left\{\beta^\top \left(X^{j_1}-X^{j_2}\right) \leq 2 W\right\},
\end{eqnarray*}
where $W = \kappa_4 \|\hat{\beta}-\beta\|_2$ is obtained by applying Assumption \ref{a1}(iii). Note that $W$ is independent of the difference $\beta^\top \left(X^{j_1}-X^{j_2}\right)$, because $W$ is computed from the historical data $\bm{H}_n$, which is independent of the new data $\left(\bm{X},\bm{Z}\right)$.

We then establish the following tail bound on the estimation error of $\hat{\beta}$. The argument for the bound is quite technical and not central to what follows, so we state the result here and give the proof in the Appendix.

\begin{thm}\label{thm:regconc}
There exist constants $C_1,C_2$ depending only on $\kappa_2,\kappa_3,\kappa_4$ such that
\begin{equation}\label{eq:regconc}
P\left(\sqrt{n}\|\hat{\beta}-\beta\|_2 > s\right) \leq C_1 e^{-C_2 s^2} + C_1e^{-C_2n}, \qquad s>0.
\end{equation}
\end{thm}

Using Theorem \ref{thm:regconc}, we can find constants $C_3,C_4$ such that
\begin{eqnarray}
&\,& P\left(\arg\max_{k:Z^k=z} \hat{\beta}^\top X^k \neq \arg\max_{k:Z^k=z}\beta^\top X^k\mid \bm{X},\bm{Z}\right)\nonumber\\
&\leq & P\left(\beta^\top \left(X^{j_1}-X^{j_2}\right) \leq 2 W\mid \bm{X},\bm{Z}\right)\nonumber\\
&=& P\left(\sqrt{n}\left(\beta^\top \left(X^{j_1}-X^{j_2}\right)\right) \leq 2\sqrt{n}\cdot W\mid \bm{X},\bm{Z}\right)\nonumber\\
&\leq& C_3 \exp\left(-C_4 n\min\left\{\frac{1}{2\kappa^2_4}\left(\beta^\top \left(X^{j_1}-X^{j_2}\right)\right)^2,1\right\}\right)\nonumber\\
&\leq& C_3 \exp\left(-\frac{1}{2\kappa^2_4} C_4 n \left(\beta^\top \left(X^{j_1}-X^{j_2}\right)\right)^2\right),\label{eq:withingroupbound}
\end{eqnarray}
where the last line follows by Assumption \ref{a1}(iii). This completes the proof of the conditional bound.

To obtain the unconditional bound, we first observe that $\left(F^{-1}_z\right)' = \left(F'_z\circ F^{-1}_z\right)^{-1}$. By Assumption \ref{a1}(v), it follows that $\left(F^{-1}_z\right)' \geq \kappa^{-1}_6$. Consequently, (\ref{eq:withingroupbound}) yields
\begin{eqnarray*}
&\,& P\left(\arg\max_{k:Z^k=z} \hat{\beta}^\top X^k \neq \arg\max_{k:Z^k=z}\beta^\top X^k\mid \bm{X},\bm{Z}\right)\\
&\leq & C_3\exp\left(-\frac{1}{2\kappa^2_4\kappa^2_6}C_4n\left(F_z\left(\beta^\top X^{j_1}\right)-F_z\left(\beta^\top X^{j_2}\right)\right)^2\right).
\end{eqnarray*}
The quantity $F_z\left(\beta^\top X^{j_1}\right)-F_z\left(\beta^\top X^{j_2}\right)$ is precisely the difference between the top two uniform order statistics from a sample of size $K_z$, and therefore follows the distribution $Beta\left(1,K_z\right)$. Because $K_z\leq K$, the density of this distribution is bounded by $K$ on $\left[0,1\right]$. We then derive
\begin{eqnarray*}
P\left(\arg\max_{k:Z^k=z} \hat{\beta}^\top X^k \neq \arg\max_{k:Z^k=z}\beta^\top X^k\right) &\leq & C_3 K \int^1_0 \exp\left(-\frac{C_4}{2\kappa^2_4\kappa^2_6}n u^2\right)du\\
&=& \frac{C_3 K \kappa_4 \kappa_6}{\sqrt{C_4 n}}\int^{\frac{\sqrt{C_4 n}}{\kappa_4 \kappa_6}}_0 \exp\left(-\frac{1}{2}z^2\right)dz\\
&\leq & \frac{C_3 K \kappa_4 \kappa_6}{\sqrt{C_4 n}}\int^{\infty}_{-\infty} \exp\left(-\frac{1}{2}z^2\right)dz\\
&=& C_5 n^{-\frac{1}{2}}
\end{eqnarray*}
for some $C_5>0$. This completes the proof of the unconditional bound.

\subsection{Bound on subgroup selection error}

It remains to bound the last term on the right-hand side of (\ref{eq:twotypes}). We require a concentration inequality on $\hat{q}$, similar to what was obtained for $\hat{\beta}$ in Theorem \ref{thm:regconc}. This quantile, however, is obtained from $\hat{T}$, an estimator of the cdf of $R^{(1)}-R^{(0)}$ which itself is derived from the empirical cdfs $F_z$. Thus, we successively derive exponential bounds on all of these quantities. As in Section \ref{sec:estimationbound}, the proofs of these results are given in the Appendix.

\begin{thm}\label{thm:cdfconc}
There exist constants $C_6,C_7,C_8 > 0$ depending only on $\kappa_1$ and $p$ such that
\begin{equation}\label{eq:cdfconc}
P\left(\sqrt{n} \sup_{z,\theta} \|\hat{F}_{z,\theta} - F_{z,\theta}\|_{\infty} > t + C_6\right) \leq C_7 e^{-C_8 t^2}, \qquad t\in\left(0,\frac{\sqrt{n}}{2\kappa_1}\right),
\end{equation}
where $\hat{F}_{z,\theta}$ is identical to $\hat{F}_z$ but with $\hat{\beta}$ replaced by the fixed vector $\theta$.
\end{thm}

Recall from (\ref{eq:T}) and (\ref{eq:That}) that $T,\hat{T}$ depend on $K_0,K_1$, which are derived from $\bm{Z}$. We prove the following results for $K_0,K_1$ fixed, but since the bounds that we obtain do not depend on these quantities, this makes no difference.

\begin{thm}\label{thm:Tconc}
Let $K_0,K_1 \geq 1$ be fixed integers satisfying $K_0+K_1 = K$, and define $T,\hat{T}$ as in (\ref{eq:T}) and (\ref{eq:That}). There exist constants $C_9,C_{10},C_{11} > 0$ depending only on $\kappa_2$, $\kappa_3$, $\kappa_4$, $p$, and $K$ such that, for any $t > 0$,
\begin{equation}\label{eq:Tconc}
P\left(\sqrt{n}\| \hat{T}-T\|_{\infty} > C_9 t + C_{10}\right) \leq 2\exp\left(-C_{11} t^2\right) + 2\exp\left(-C_{11}n\right).
\end{equation}
\end{thm}

\begin{thm}\label{thm:qconc}
Let $K_0,K_1 \geq 1$ be fixed integers satisfying $K_0+K_1 = K$, and let $q,\hat{q}$ satisfy $T\left(q\right) = \hat{T}\left(\hat{q}\right) = 0$. There exist constants $C_{12},C_{13}>0$ depending only on $\kappa_2$, $\kappa_3$, $\kappa_4$, $p$, and $K$ such that, for any $t > 0$,
\begin{equation}\label{eq:qconc}
P\left(\sqrt{n}\left|\hat{q}-q\right| > C_{12}t\right) \leq 2\exp\left(-C_{13} t^2\right) + 2\exp\left(-C_{13}n\right).
\end{equation}
\end{thm}

We now derive the conditional bound. Note that $q,K_0,K_1$ can be treated as fixed when $\left(\bm{X},\bm{Z}\right)$ is given. Intuitively, we will select the incorrect subgroup if the estimation error in $\hat{q},\hat{\beta}$ is greater in some sense than the gap $\left|R^{(1)}-R^{(0)}-q\right|$. Formally, we write
\begin{eqnarray}
&\,& P\left(1_{\left\{\hat{R}^{(1)}- \hat{R}^{(0)}\geq \hat{q}\right\}}\neq 1_{\left\{R^{(1)}- R^{(0)}\geq q\right\}}\mid \bm{X},\bm{Z}\right)\nonumber\\
&\leq& P\left( \left|R^{(1)}-R^{(0)}-q\right| \leq \left|\hat{q}-q\right|+2\kappa_4\|\hat{\beta}-\beta\|_2\mid \bm{X},\bm{Z}\right)\nonumber\\
&\leq& P\left(\sqrt{n}\left|\hat{q}-q\right|\geq \frac{\sqrt{n}}{2}\left|R^{(1)}-R^{(0)}-q\right|\mid\bm{X},\bm{Z}\right)\nonumber\\
&\,& + P\left(\sqrt{n}\|\hat{\beta}-\beta\|_2 \geq \frac{\sqrt{n}}{4\kappa_4}\left|R^{(1)}-R^{(0)}-q\right|\mid\bm{X},\bm{Z}\right).\label{eq:splitgroupselection}
\end{eqnarray}
The estimation errors $\hat{\beta}-\beta$ and $\hat{q}-q$ are independent of $\left(\bm{X},\bm{Z}\right)$, so we can apply Theorem \ref{thm:regconc} to obtain
\begin{eqnarray}
&\,& P\left(\sqrt{n}\|\hat{\beta}-\beta\|_2 \geq \frac{\sqrt{n}}{4\kappa_4}\left|R^{(1)}-R^{(0)}-q\right|\mid\bm{X},\bm{Z}\right)\nonumber\\
&\leq & 4C_1\exp\left(-\frac{C_2 n}{16\kappa^2_4}\min\left\{\left(R^{(1)}-R^{(0)}-q\right)^2,16\kappa^2_4\right\}\right)\nonumber\\
&=& 4C_1\exp\left(-\frac{C_2 n}{16\kappa^2_4}\left(R^{(1)}-R^{(0)}-q\right)^2\right),\label{eq:splitterm1}
\end{eqnarray}
where the last equality follows by Assumption \ref{a1}(iii), which also implies $\left|q\right|\leq 2\kappa_4$.

Similarly, we apply Theorem \ref{thm:qconc} to obtain
\begin{eqnarray}
P\left(\sqrt{n}\left|\hat{q}-q\right|\geq \frac{\sqrt{n}}{2}\left|R^{(1)}-R^{(0)}-q\right|\mid\bm{X},\bm{Z}\right) &\leq & 4\exp\left(-\frac{1}{4}C_{13}n\min\left\{\left(R^{(1)}-R^{(0)}-q\right)^2,4\right\}\right)\nonumber\\
&\leq& 4\exp\left(-\frac{1}{4}C_{13}C_{14}n\left(R^{(1)}-R^{(0)}-q\right)^2\right),\label{eq:splitterm2}
\end{eqnarray}
where $C_{14} = \min\left\{1,\frac{1}{4}\kappa^{-2}_4\right\}$. Combining (\ref{eq:splitgroupselection}) with (\ref{eq:splitterm1})-(\ref{eq:splitterm2}), we obtain
\begin{equation}\label{eq:splitfinal}
P\left(1_{\left\{\hat{R}^{(1)}- \hat{R}^{(0)}\geq \hat{q}\right\}}\neq 1_{\left\{R^{(1)}- R^{(0)}\geq q\right\}}\mid \bm{X},\bm{Z}\right) \leq C_{16}\exp\left(-C_{15} n\left(R^{(1)}-R^{(0)}-q\right)^2\right),
\end{equation}
where $C_{15} = \min\left\{\frac{C_2}{16\kappa^2_4},\frac{1}{4}C_{13}C_{14}\right\}$ and $C_{16}=8\left(C_1+1\right)$. This completes the proof of the conditional bound.

We now turn to the unconditional bound. First, let us write
\begin{eqnarray*}
\frac{d}{dt} P\left(R^{(1)}-R^{(0)} \leq t\mid\bm{Z}\right) &=& \frac{d}{dt}\int G_{1,\beta}\left(u+t\right)dG_{0,\beta}\left(u\right)\\
&=& \int K_1\left(F_{1,\beta}\left(u+t\right)\right)^{K_1-1}F'_{1,\beta}\left(u+t\right)dG_{0,\beta}\left(u\right),
\end{eqnarray*}
using the fact that $G_{1,\beta} = \left(F_{1,\beta}\right)^{K_1}$. The interchange of differentiation and integration at $t=q$ is valid by Thm. 2.27 of \cite{Fo99} because the density $F'_{z,\beta}$ is bounded by $\kappa_6$ due to Assumption \ref{a1}(v). By the same assumption, the conditional density of $R^{(1)}-R^{(0)}$ given $\bm{Z}$ is bounded by $K\kappa_6$. Applying this bound to (\ref{eq:splitfinal}), and using arguments similar to those at the end of Section \ref{sec:estimationbound}, yields the unconditional bound.

\section{Computational experiments}\label{sec:experiments}

We present numerical experiments demonstrating the practical potential of our approach for making decisions that are both optimal and fair. Section \ref{sec:benchmark} describes two benchmarks, based on regularized estimation, from the literature on fair prediction, which were also implemented in the experiments. Section \ref{sec:synthetic} considers a setting with synthetic (simulated) data. Section \ref{sec:real} presents results using real data on law students; separate experiments are conducted with two different protected attributes.

\subsection{Benchmark policies}\label{sec:benchmark}

Many of the models and procedures in the extensive literature on fair prediction were designed for classification problems, where the response variable $Y$ is binary. Though our approach can potentially handle such problems as well (since it does not impose much structure on the cdf $F_z$ connecting the linear model to the response), our primary focus in this paper is on linear regression. Among those papers that are able to handle this setting, the predominant method of computing fair predictions is to solve a modified version of the usual statistical estimation problem (choosing regression coefficients that minimize total statistical error across the given dataset) with additional constraints that impose a bound on some measure of unfairness, such as the correlation between the prediction and the protected attribute. One can also reformulate such constrained optimization problems by turning the constraint into a regularization term; in the context of linear regression, one thus solves
\begin{equation*}
\min_\theta \|\bm{Y}_n - \bm{X}_n\theta\|_2 + \lambda L\left(\bm{H}_n,\theta\right),
\end{equation*}
where $L$ is the chosen measure of unfairness, and $\lambda \geq 0$ is a tunable parameter governing the tradeoff between fairness and accuracy. In the hiring context, having obtained an estimator $\hat{\theta}$ from the regularized problem, our decision is given by $\arg\max_k \hat{\theta}^\top X^k$.

Many penalties have been proposed in the literature and evaluated empirically on various datasets. We choose two penalties from recent papers that we believe to be representative of this class of methods. The first penalty, proposed by \cite{Berk17}, is written as
\begin{equation*}
L\left(\bm{H}_n,\theta\right) = \frac{\sum_{Z_m=1,Z_{m'}=0}w\left(Y_m,Y_{m'}\right)\left(\theta^\top X_m - \theta^\top X_{m'}\right)^2}{\left(\sum^n_{m=1}1_{\left\{Z_m=1\right\}}\right)\left(\sum^n_{m=1}1_{\left\{Z_m=0\right\}}\right)},
\end{equation*}
where the authors recommend to set the weights $w\left(y,y'\right) = e^{-\left(y-y'\right)^2}$. The goal of this method is to make similar predictions for similar response variables, for either group. The second penalty, considered by \cite{ZiRo20}, has the simpler form
\begin{equation*}
L\left(\bm{H}_n,\theta\right) = \left(\frac{\sum_{Z_m=1} Y_m-\theta^\top X_m}{\sum^n_{m=1}1_{\left\{Z_m=1\right\}}} - \frac{\sum_{Z_m=0} Y_m-\theta^\top X_m}{\sum^n_{m=1}1_{\left\{Z_m=0\right\}}}\right)^2,
\end{equation*}
which seeks to balance the prediction error between the two groups. Both papers also consider other modified versions of these penalties, and still others can be found in other papers, but there is much similarity between them (for example, every such approach requires a tunable parameter), and no consensus on which specific version is the best. In fact, as will become clear later in this section, both of the above penalties lead to very similar performance as far as \textit{decision-making} is concerned, which suggests that the nuances between penalties are not the central issue in our problem.

An important practical issue, which we encountered in our experiments, but which does not seem to have attracted much attention in the literature, is the computational cost of solving the penalized regression problem. Both \cite{Berk17} and \cite{ZiRo20} used the CVX convex optimization package \citep{cvx} to solve their respective problems. We did this as well, and found that the computation time was non-negligible even for a single problem, to say nothing of the multiple problems one has to solve in order to tune $\lambda$. In particular, the penalty of \cite{Berk17} adds up a large number of terms, and thus is time-consuming to evaluate even for a fairly small sample size of $n=1000$. We reduced this cost by using sparse matrices and setting the weights $w\left(y,y'\right)$ to zero whenever the recommended distance $e^{-\left(y-y'\right)^2} < 10^{-5}$. Even so, the resulting problem was significantly slower to solve than that of \cite{ZiRo20}. By contrast, our proposed policy ran much faster. To give an example, in the setting of Section \ref{sec:synthetic}, the time required to compute a single decision for a sample size of $n=1000$ was approximately $1.25$s for the method of \cite{Berk17}, $0.25$s for the method of \cite{ZiRo20}, and $0.001$s for our method. Our experiments were conducted in MATLAB, but we used the interface between CVX and the commercial solver MOSEK to solve the penalized problems; this was much faster than CVX's default solver, but still much slower than our method.

\subsection{Synthetic data}\label{sec:synthetic}

We present one synthetic scenario that cleanly illustrates how our approach can address certain important practical concerns, with the caveat that many of the specifics of actual hiring decisions are abstracted away. First, let $\mathbb{E}\left(Z\right) = 0.15$. Let $p = 30$ be the dimensionality of the problem. Given $Z = z$, we assume that $X\sim\mathcal{N}\left(0,\bm{C}_z\right)$, where the $p\times p$ covariance matrix is chosen according to $\bm{C}_z = \tau_z \bm{A}_z\bm{A}^\top_z$, where $\bm{A}_z$ is a matrix of independent standard normal random variables, and
\begin{equation*}
\tau_z = \left\{
\begin{array}{c c}
\frac{1}{2} & z = 1,\\
1 & z = 0.
\end{array}
\right.
\end{equation*}
As a result, realizations of $X$ given $Z = 1$ tend to have somewhat lower spread than given $Z = 0$, even though the conditional mean of the performance $\beta^\top X$ is the same in both cases.\footnote[3]{It bears repeating that this difference between distributions is the \textit{effect} of numerous structural and systemic factors that are not directly visible to the decision-maker.} Since we are choosing the \textit{best} among $K$ candidates, however, this means that parity of treatment (as discussed in Section \ref{sec:noparity}) will underrepresent the minority group.

We generated the historical data $\left(\bm{X}_n,\bm{Z}_n\right)$ from the above-described distributions with a sample size of $n = 1000$. A single vector $\beta$ was generated from a $\mathcal{N}\left(0,\bm{I}_p\right)$ distribution and fixed at the beginning of the experiment; the covariance matrices $\bm{C}_z$ were also generated once and then fixed. The historical responses $\bm{Y}_n$ were calculated according to (\ref{eq:reg}) with i.i.d. standard normal residual errors. The new data $\left(\bm{X},\bm{Z}\right)$ used for decision-making were also sampled from the data-generating distribution with $K = 10$ candidates per decision.

In our experiments, the empirical policy $\hat{\pi}$ from Section \ref{sec:empirical} is shown the first $m \leq n$ of the historical data points, uses these data to compute the relevant estimators, then selects from among a new independently generated set of $K$ candidates. The precise sequence of selection decisions does not influence the historical data; the purpose of varying $m$ is simply to show the effect of sample size on performance. The performance metrics are the objective value $\mathbb{E}\left(\beta^\top X^{\hat{\pi}}\right)$, as well as the probability $P\left(Z^{\hat{\pi}}=1\right)$ of hiring a minority candidate, both estimated empirically by averaging over $10,000$ macro-replications at each $m$ value (the large number of macro-replications renders the standard errors negligible). We also implemented parity of treatment, given by $\arg\max_k \hat{\beta}^\top X^k$, in the same manner.

\begin{figure}[t]
	\centering
	\subfigure[Performance over time.]{
		\includegraphics[width=0.47\textwidth]{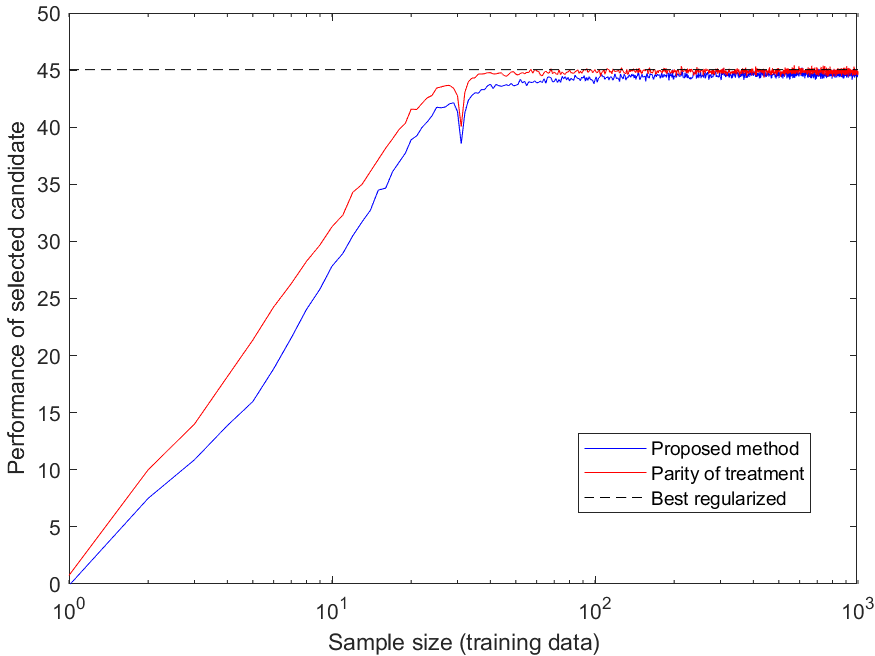}
		\label{fig:synth1}
	}
	\subfigure[Parity over time.]{
		\includegraphics[width=0.47\textwidth]{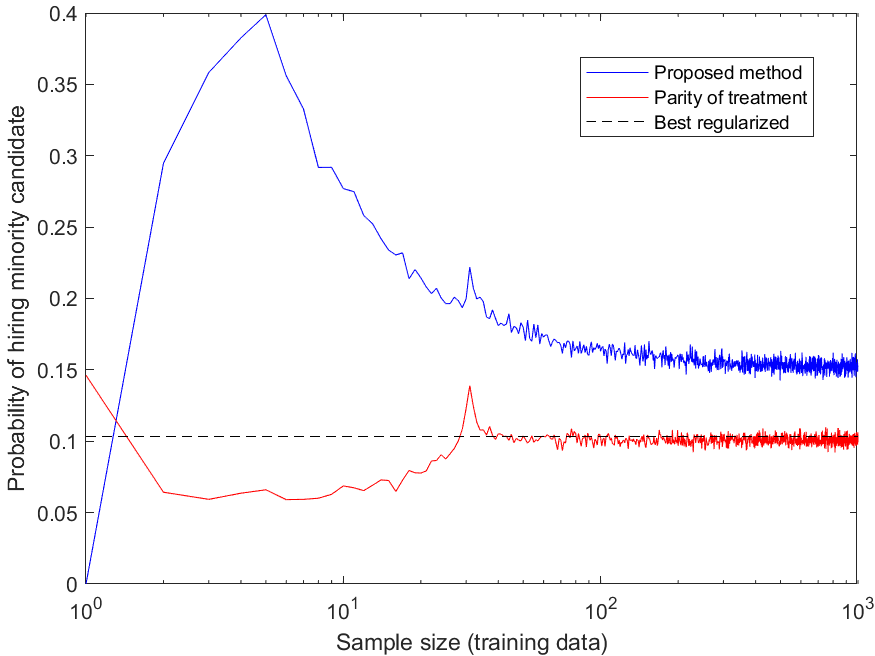}
		\label{fig:synth2}
	}
\caption{Empirical performance and parity for synthetic data.}\label{fig:synthetic}
\end{figure}

With regard to the two regularization-based methods used as benchmarks, it was not practically feasible to run them for every $m$ value (with multiple macro-replications) due to the high computational cost of running convex programming solvers. Thus, we only ran them for the entire sample size of $1000$ data points and reported the results (averaged over $100$ macro-replications) for the best of the two as horizontal lines in Figure \ref{fig:synthetic}. In this way, we do not see how the performance of these methods varies with sample size, but we can see their ``best'' performance with the largest possible sample size and compare this to the trajectories of the other two methods.

In Figure \ref{fig:synth1}, we see that both parity of treatment and the proposed empirical fair policy show improved performance over time as $m$ increases and $\beta$ is estimated more accurately. Unsurprisingly, the fair policy lags slightly behind during this period, because parity of treatment is able to maximize reward without any constraints, and because our policy also has to estimate a difficult cdf in addition to $\beta$. What is remarkable, however, is that this gap closes almost completely for larger sample sizes: at $m = 1000$, the objective value achieved by the fair policy is $99.76\%$ of the value achieved by the unfair one! At the same time, Figure \ref{fig:synth2} shows that the unfair policy hires significantly fewer minority candidates than there are in the candidate population ($9.92\%$ at $m = 1000$), while the fair policy has achieved statistical parity.

It is interesting to observe that neither of the regularization-based benchmarks was able to improve over parity of treatment to any appreciable degree. We ran both methods over a wide range of magnitudes of the regularization parameter $\lambda$, but found that the results (shown in Table \ref{tab:lambda}) were both remarkably insensitive and very similar between the two methods. For this reason, only the best performance values in the table are shown in Figure \ref{fig:synthetic}.

The ineffectiveness of the regularization-based methods highlights the distinction between fair prediction and fair decision-making. Both of these methods minimize an adjusted form of prediction error in order to reduce bias in the prediction: for example, a proposed vector $\theta$ of regression coefficients may lead to bias if the random variable $\left(\beta-\theta\right)^\top X$ is correlated with $Z$. In our setting, however, it is possible to have \textit{no} prediction error (i.e., $\theta = \beta$) and still make an unfair decision. Furthermore, if we have $K$ candidates to choose from, the prediction error for most of these candidates is of little importance; if we incorrectly rank some candidates in the middle or bottom of the pool, this in itself does not matter much. The key issue is how well we can distinguish between candidates on the tails of the distributions of their respective groups. Because regularization-based formulations do not have any notion of decision-making built in, they turn out to show no tangible improvement over parity of treatment.

\begin{table}[t] 
	\centering
	\begin{tabular}{|c||c|c||c|c|}
		\hline
        \multirow{2}{*}{$\lambda$} & \multicolumn{2}{|c||}{\cite{Berk17}} & \multicolumn{2}{|c|}{\cite{ZiRo20}}\\ \cline{2-5}
         & Performance & Parity & Performance & Parity \\ \hline
         $10^{-4}$	& 44.9381 &	0.1014 &	45.0064 &	0.0985\\
        $10^{-3}$ &	45.0441 &	0.1015 &	44.9409 &	0.1013\\
        $10^{-2}$ &	44.8315 &	0.1015 &	44.9085 &	0.1001\\
        $10^{-1}$ &	44.8547 &	0.0998 &	44.8405 &	0.1033\\
        $10^{0}$ &	44.8811 &	0.1013 &	44.9203 &	0.1012\\
        $10^1$ &	44.9478 &	0.1019 &	44.9457 &	0.1002\\
        $10^2$ &	44.7895 &	0.1020 &	44.8667 &	0.1011\\
        $10^3$ &	44.8426 &	0.1018 &	44.8341 &	0.1000\\
        $10^4$ &	44.9804 &	0.0987 &	44.9362 &	0.1010\\
        \hline
	\end{tabular}
	\caption{Performance of regularization-based benchmarks as a function of $\lambda$.\label{tab:lambda}}
\end{table}

From this example, we see that fairness does not have to come at the expense of performance. As explained in Section \ref{sec:idealprop}, the ideal fair policy favors minority candidates in those situations where they are very strong relative to non-minority candidates (i.e., the best minority candidate is strong, or the best non-minority candidate is weak). In this way, the employer's interests are taken into account while guaranteeing equity for the underrepresented group.

\subsection{Real data}\label{sec:real}

We also evaluated our approach on real data\footnote[4]{Publicly available at: \url{http://www.seaphe.org/databases.php}} from the LSAC National Longitudinal Bar Passage Study \citep{Wi98}. The dataset describes (in anonymized form) over $27,000$ students from various US law schools. For our purposes, we let $X$ consist of those attributes that could be observed when the students first applied (family income, LSAT score, undergraduate GPA, birth year, and whether they were applying to a part-time or full-time program), and we used students' final GPA in law school as the response variable $Y$. As described in \cite{Wi98}, the original study had already adjusted final GPA to correct for differences between schools. We considered two possible choices for the protected attribute $Z$, namely gender (female) and race (Black/Hispanic). In the following, we ran a separate set of experiments for each of these two attributes. After removing entries with missing data, we were left with $N = 22,337$ students.

We do not know the true distribution of the population that was sampled in this study. For the purpose of these experiments, we chose to treat the sample of $N$ students as the entire population. Rather than fit a vector $\beta$ to the population, we use the observed response value $Y$ of each candidate to measure performance. We also treat the empirical distribution of $\left(X,Z\right)$ over the population of $N$ students as the data-generating process from which a smaller sample of size $n$ can be drawn and used to train different methods. Essentially, the history $\bm{H}_n$ and the new data $\left(\bm{X},\bm{Z}\right)$ are all being bootstrapped from the original dataset.

We used $K = 30$ in both experiments. The proportion $\mathbb{E}\left(Z\right)$ was $0.4383$ for women, and $0.0642$ for Black/Hispanic applicants. In this way, our method can be tested on settings whose respective proportions of minority candidates are relatively high and low. Overall, we followed the same approach for implementing and evaluating methods as in Section \ref{sec:synthetic}.

\begin{figure}[t]
	\centering
	\subfigure[Performance over time.]{
		\includegraphics[width=0.47\textwidth]{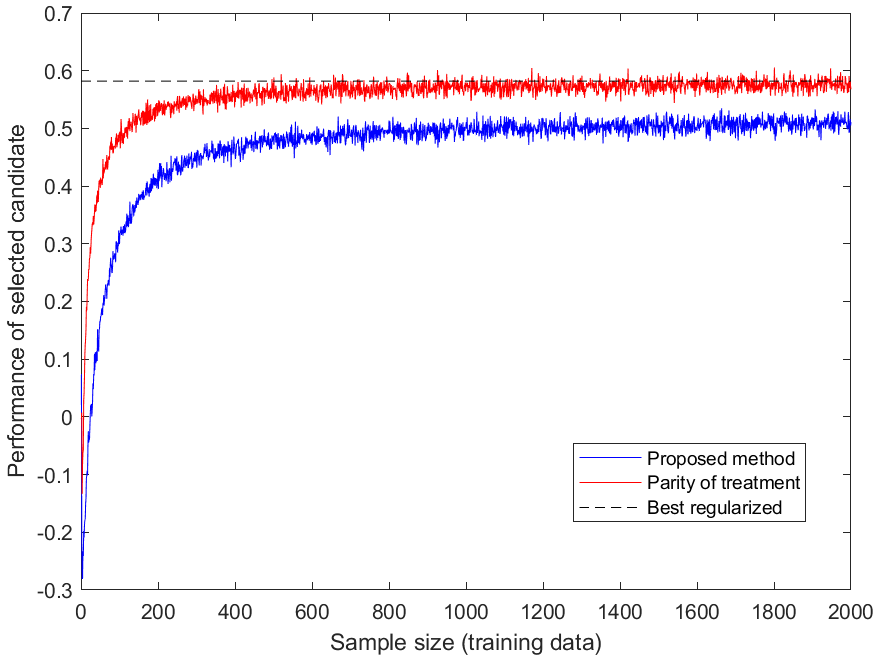}
		\label{fig:lsacrace1}
	}
	\subfigure[Parity over time.]{
		\includegraphics[width=0.47\textwidth]{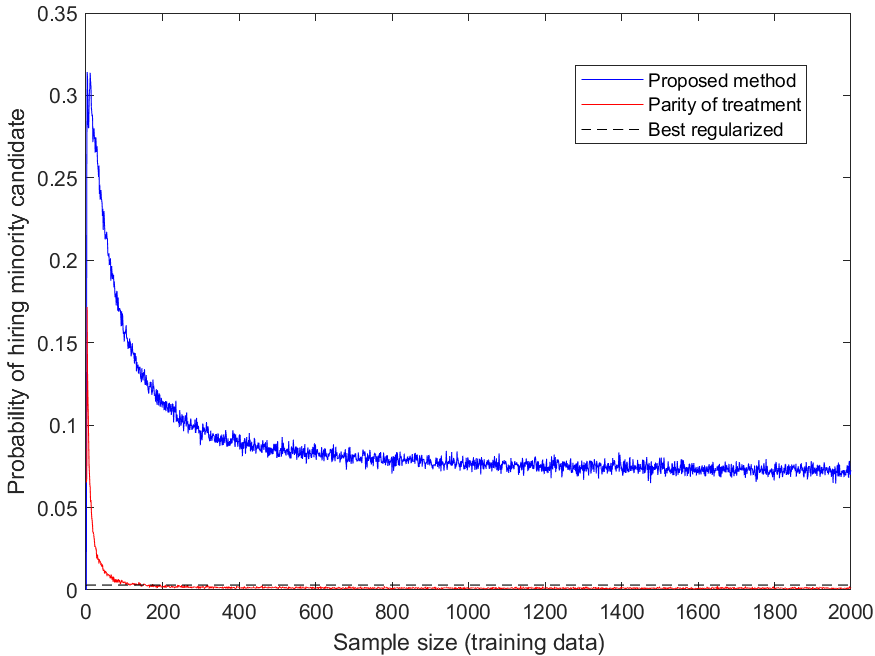}
		\label{fig:lsacrace2}
	}
\caption{Empirical performance and parity with race as the protected attribute.}\label{fig:lsacrace}
\end{figure}

Figure \ref{fig:lsacrace} presents results for the setting where race was used as the protected attribute, with $n = 2000$ as the largest sample size. The data reflect a significant disparity between subgroups, with $\mathbb{E}\left(Y\mid Z=1\right)-\mathbb{E}\left(Y\mid Z=0\right)=-1.0809$ (again, the expectation here is over the empirical distribution in the data). Nonetheless, as seen in Figure \ref{fig:lsacrace1}, the gap between the objective values of the fair policy and the unfair parity of treatment policy has a much smaller magnitude of $0.05$. The fair policy achieves $90\%$ of the objective value of the unfair policy. At the same time, the difference in equity between the two policies (Figure \ref{fig:lsacrace2}) is particularly striking: all of the benchmarks\footnote[5]{Again, we ran both regularization-based methods for a range of $\lambda$ values, and reported the best performance across all cases in Figure \ref{fig:lsacrace}. We do not repeat Table \ref{tab:lambda} again as the numbers are not particularly informative.} virtually never hire minority candidates. Only the proposed policy is able to mitigate this problem.

\begin{figure}[t]
	\centering
	\subfigure[Performance over time.]{
		\includegraphics[width=0.47\textwidth]{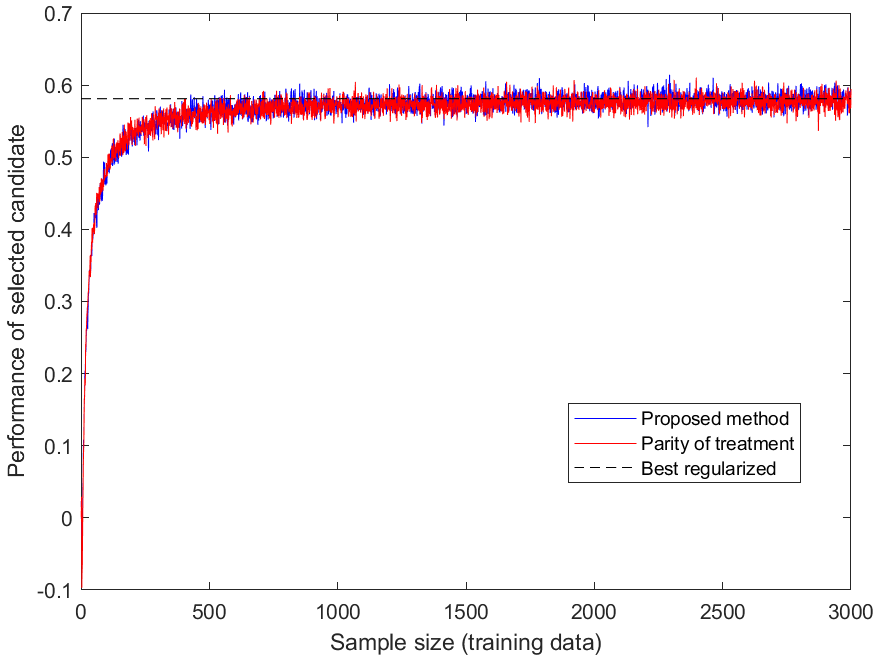}
		\label{fig:lsacgender1}
	}
	\subfigure[Parity over time.]{
		\includegraphics[width=0.47\textwidth]{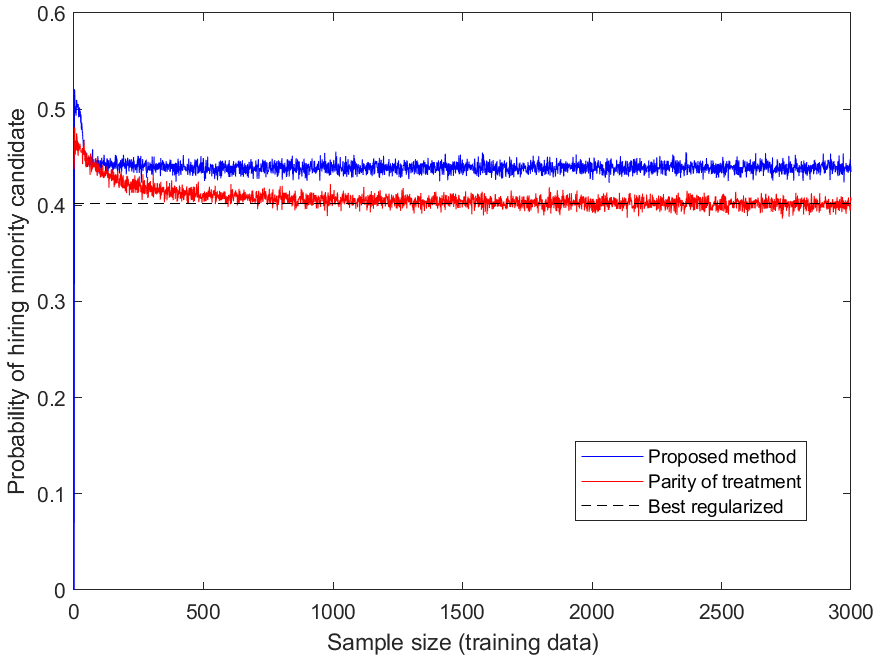}
		\label{fig:lsacgender2}
	}
\caption{Empirical performance and parity with gender as the protected attribute.}\label{fig:lsacgender}
\end{figure}

Figure \ref{fig:lsacgender} presents analogous results with gender as the protected attribute and $n = 3000$ as the largest sample size. This time, the data reflect a smaller disparity between subgroups, with $\mathbb{E}\left(Y\mid Z=1\right)-\mathbb{E}\left(Y\mid Z=0\right)\approx 0$. Consequently, as shown in Figure \ref{fig:lsacgender1}, the fair policy incurs virtually no loss in objective value relative to the unfair policy. However, this does mean that there is no fairness issue: as seen in Proposition \ref{prop:extreme}, the hiring decision is influenced by the tails of the conditional distributions, not by their means. We see in Figure \ref{fig:lsacgender2} that the unfair policy underrepresents women, who comprise $43.84\%$ of this population but are selected by the policy only $40\%$ of the time.\footnote[6]{Regarding the regularization-based methods, we could only run the penalty of \cite{ZiRo20} when gender was the protected attribute. The approach of \cite{Berk17} did not scale, with over $2$ million terms in the objective function even after the tweaks described in Section \ref{sec:benchmark}.} The fair policy completely corrects this disparity.

To summarize, the proposed policy achieves statistical parity, with realistic sample sizes, across all examples, which include both synthetic and real data, as well as varied frequency of the protected attribute (as low as $0.0642$ and as high as $0.4383$). A major insight of these experiments is that the firm can ensure fairness with only minor reduction in objective value, and often (as shown in Figures \ref{fig:synth1} and \ref{fig:lsacgender1}) with virtually no reduction at all.

\section{Conclusion}

We have presented a data-driven prescriptive framework for fair decisions in a stylized hiring problem where applicants are drawn independently from a population, and the firm evaluates their expected performance based on their observable attributes. Our main contribution is a hiring policy that is both fair with respect to a protected attribute of the candidates, and optimal (among all possible fair policies) with respect to the firm's objective of hiring the best talent. Crucially, this policy does \textit{not} ignore the protected attribute; on the contrary, it depends on the attribute \textit{functionally} in order to ensure that the decision does not depend on it \textit{probabilistically}. The policy is transparent, interpretable, and easy to estimate empirically in a provably optimal manner. Our numerical results show that the policy is highly practical and performs well even with fairly small sample sizes. Furthermore, in many situations, fairness is achievable with negligible effect on objective value -- that is, decisions can be made much less damaging to minority candidates at very little cost to the firm.

Our work contributes to the growing literature on data-driven decision-making, which aims to integrate ideas from statistics and machine learning into OR/OM models rather than simply applying ML algorithms out of the box. In particular, we offer a new perspective on how fairness may be studied within a decision problem, and we show that fair decisions are very different from fair predictions. The presence of an economic objective function guides the selection of a fair policy and helps obtain decisions that meet firms' expectations while reducing the destructive effects of bias. We view fair decision-making as a rich problem class, separate from fair prediction, and we envision many possible extensions of the ideas presented in this paper to other classes of decision problems in revenue management, service operations and other areas.


\bibliographystyle{apalike}
\bibliography{fairness} 

\section{Appendix: exact computation of the empirical quantile}\label{sec:appendixquantile}

Recall from Section \ref{sec:upperbound} that the empirical quantile $\hat{q}$ used in the computation of $\hat{\pi}$ is the smallest value satisfying $\hat{T}\left(\hat{q}\right) \geq 0$, where
\begin{equation}\label{eq:newThat}
\hat{T}\left(t\right) = \frac{1}{n_0}\sum^n_{m=1} \left(\hat{F}_1\left(\hat{\beta}^\top X_m + t\right)\right)^{K_1}\left(\hat{F}_0\left(\hat{\beta}^\top X_m\right)^{K_0-1}\right) 1_{\left\{Z_m=0\right\}} - \frac{1}{K},
\end{equation}
with
\begin{equation*}
n_z = \sum^n_{m=1} 1_{\left\{Z^m=z\right\}}.
\end{equation*}
In the following, we show how $\hat{q}$ can be computed exactly without the need for bootstrapping. Again, we treat $\hat{\beta}$, $\bm{H}_n$, $\bm{Z}$ and $K_0,K_1$ as fixed, since these are known at the time when we compute $\hat{\pi}$.

First, let $\ell^{(z)}_1,...,\ell^{(z)}_{n_z}$ be the elements of the set $\left\{m\leq n\,:\,Z^m=z\right\}$ chosen in such a way that
\begin{eqnarray*}
\hat{\beta}^\top X_{\ell^{(1)}_1} \geq \hat{\beta}^\top X_{\ell^{(1)}_2} \geq ... \geq \hat{\beta}^\top X_{\ell^{(1)}_{n_1}},\\
\hat{\beta}^\top X_{\ell^{(0)}_1} \leq \hat{\beta}^\top X_{\ell^{(0)}_2} \leq ... \leq \hat{\beta}^\top X_{\ell^{(0)}_{n_0}}.
\end{eqnarray*}
Then, (\ref{eq:newThat}) can be rewritten as
\begin{equation*}
\hat{T}\left(t\right) = \frac{1}{n_0}\sum^{n_0}_{m=1} \left(\hat{F}_1\left(\hat{\beta}^\top X_{\ell^{(0)}_m} + t\right)\right)^{K_1}\left(\frac{m}{n_0}\right)^{K_0-1} - \frac{1}{K},
\end{equation*}
because the values $\hat{\beta}^\top X_{\ell^{(0)}_m}$ are increasing. Recall also that
\begin{equation*}
\hat{F}_1\left(\hat{\beta}^\top X_{\ell^{(0)}_m}+t\right) = \frac{1}{n_1}\sum^{n_1}_{m'=1} 1_{\left\{\hat{\beta}^\top\left(X_{\ell^{(1)}_{m'}}-X_{\ell^{(0)}_m}\right) \leq t\right\}}.
\end{equation*}

Then, the computation can proceed as follows. Define a matrix $B\in\mathds{R}^{n_1\times n_0}$ with $B_{m',m} = \hat{\beta}^\top\left(X_{\ell^{(1)}_{m'}}-X_{\ell^{(0)}_m}\right)$. Then, $B_{m',m}$ decreases in both $m'$ and $m$. Consequently,
\begin{equation*}
\hat{F}_1\left(\hat{\beta}^\top X_{\ell^{(0)}_m}+t\right) = \frac{1}{n_1}\sum^{n_1}_{m'=1} 1_{\left\{B_{m',m}\leq t\right\}}.
\end{equation*}
Then,
\begin{equation*}
\hat{T}\left(t\right) = \frac{1}{n_0}\sum^{n_0}_{m=1} \left(\frac{1}{n_1}\sum^{n_1}_{m'=1} 1_{\left\{\hat{\beta}^\top\left(X_{\ell^{(1)}_{m'}}-X_{\ell^{(0)}_m}\right) \leq t\right\}}\right)^{K_1}\left(\frac{m}{n_0}\right)^{K_0-1} - \frac{1}{K}.
\end{equation*}
Since we are looking for an empirical quantile, it is sufficient to find the smallest $B_{m',m}$ value such that $\hat{T}\left(B_{m',m}\right)\geq 0$. Note that $\hat{T}$ is increasing in $t$, and by the structure of $B$, $\hat{T}\left(B_{m',m}\right)\geq 0$ implies that $\hat{T}\left(B_{j',j}\right) \geq 0$ for $j'\leq m'$ and $j\leq m$. We then need to find $m''$ such that $\hat{T}\left(B_{m'',m}\right) \geq 0$ and $\hat{T}\left(B_{m''+1,m}\right) < 0$, whence the desired $\hat{q}$ value will be equal to $B_{m''+1,j}$ for some $j \leq m$. This can be done via a simple grid search.

\section{Appendix: proofs}

In the following, we give complete proofs for all results that were stated in the main text.

\subsection{Proof of Proposition \ref{prop:extreme}}

First, we observe that
\begin{equation*}
P\left(Z^{\pi^{\max}}=1\mid \beta^\top X^{\pi^{\max}}\right) = \eta\left(\beta^\top X^{\pi^{\max}}\right)
\end{equation*}
because the conditional distribution of $Z^{\pi^{\max}}$ given $X^{\pi^{\max}}$ is the same as the conditional distribution of $Z$ given $\beta^\top X$. Next, we note that
\begin{equation*}
P\left(\beta^\top X^{\pi^{\max}} \leq r\right) = P\left(\bigcap_k \left\{\beta^\top X^k \leq r\right\}\right) = F\left(r\right)^K,
\end{equation*}
and write
\begin{eqnarray*}
P\left(Z^{\pi^{\max}} = 1\right) &=& \mathbb{E}\left(P\left(Z^{\pi^{\max}}=1\mid \beta^\top X^{\pi^{\max}}\right)\right)\\
&=& K \int \eta\left(r\right) F\left(r\right)^{K-1}dF\left(r\right)\\
&=& K \mathbb{E}\left(\eta\left(R\right)F\left(R\right)^{K-1}\right)\\
&=& K \mathbb{E}\left(\eta\left(F^{-1}\left(U\right)\right)U^{K-1}\right),
\end{eqnarray*}
where $R$ is a random variable with cdf $F$, and $U = F\left(R\right)$ is uniform on $\left[0,1\right]$. Thus, we obtain
\begin{equation*}
P\left(Z^{\pi^{\max}} = 1\right) = K\int^1_0 \eta\left(F^{-1}\left(u\right)\right)u^{K-1}du.
\end{equation*}

Now, fix an arbitrary $0 < \varepsilon <1$ and let $q_K = \varepsilon^{\frac{1}{K}}$. We see that $q_K \nearrow 1$. Observe that
\begin{eqnarray*}
\left|P\left(Z^{\pi^{\max}} = 1\right) - K\int^1_{q_K} \eta\left(F^{-1}\left(u\right)\right)u^{K-1}du\right| &=& K\int^{q_K}_0 \eta\left(F^{-1}\left(u\right)\right)u^{K-1}du\\
&\leq & K\int^{q_K}_0 u^{K-1}du\\
&=& \varepsilon.
\end{eqnarray*}
In other words,
\begin{equation}\label{eq:extreme1}
K\int^1_{q_K} \eta\left(F^{-1}\left(u\right)\right)u^{K-1}du - \varepsilon \leq P\left(Z^{\pi^{\max}} = 1\right)\leq K\int^1_{q_K} \eta\left(F^{-1}\left(u\right)\right)u^{K-1}du+\varepsilon
\end{equation}
For sufficiently large $K$, we have $\sup_{q_K\leq u \leq 1} \eta\left(u\right) \leq \limsup_{q\nearrow 1} \eta\left(q\right) +\varepsilon$. Therefore,
\begin{eqnarray*}
K\int^1_{q_K} \eta\left(F^{-1}\left(u\right)\right)u^{K-1}du &\leq & \left(\limsup_{q\nearrow 1} \eta\left(q\right) +\varepsilon\right)K\int^1_{q_K} u^{K-1}du\\
&=& \left(\limsup_{q\nearrow 1} \eta\left(q\right) +\varepsilon\right)\left(1-\varepsilon\right).
\end{eqnarray*}
Symmetrically, for sufficiently large $K$ we also have
\begin{equation*}
K\int^1_{q_K} \eta\left(F^{-1}\left(u\right)\right)u^{K-1}du \geq \left(\liminf_{q\nearrow 1} \eta\left(q\right) -\varepsilon\right)\left(1-\varepsilon\right)
\end{equation*}
Thus, (\ref{eq:extreme1}) becomes
\begin{equation*}
\left(\liminf_{q\nearrow 1} \eta\left(q\right) -\varepsilon\right)\left(1-\varepsilon\right) - \varepsilon \leq P\left(Z^{\pi^{\max}} = 1\right) \leq \left(\limsup_{q\nearrow 1} \eta\left(q\right) +\varepsilon\right)\left(1-\varepsilon\right) + \varepsilon
\end{equation*}
Since this inequality holds for arbitrary $\varepsilon$ and all sufficiently large $K$, the desired result follows.

\subsection{Proof of Theorem \ref{thm:xandj}}

For simplicity, let $A_{01} = A\left(0,1,\bm{H}_n\right)$ and $A_{11} = A\left(1,1,\bm{H}_n\right)$, suppressing the dependence on the historical data (which can take on arbitrary values). We can replace $A_{01}$ with its interior $A^\circ_{01}$ since $A_{01}\setminus A^\circ_{01}$ contains no open sets and thus has zero probability, due to the assumption that $\bm{X}$ has a density and $P\left(\bm{X} \in A_{01}\right) > 0$. Define
\begin{equation}\label{eq:Jdef}
J = \left\{ x^2 \,:\, \left\{x^1,x^2\right\}\in A_{01} \mbox{ for some } x^1\right\}.
\end{equation}
In words, $J$ is the set of all hypothetically possible candidates (i.e., their $X$ values) with $Z=1$ for which there exists at least one possible candidate with $Z=0$ who is preferred under policy $\pi$. We will show that $A\left(0,1,\bm{H}_n\right) = \mathcal{X} \times J$ and $A\left(1,1,\bm{H}_n\right) = J \times J$, and finally conclude that $J = \mathcal{X}$. The proof proceeds in four steps.

\noindent\textit{Step 1:} We show that, for any $x\in J$, there exists an open interval $D$ such that $x\in D$ and $D\times D \subseteq A_{11}$.

Fix $x\in J$. Then, there exists $x'$ such that $A_{01}$ contains an open neighborhood of $\left(x',x\right)$. Therefore, there exist two open intervals $D',D$ such that $\left(x',x\right) \in D'\times D$ and $D'\times D \subseteq A_{01}$. Let
\begin{equation*}
g_0\left(x\right) = \frac{1_{\left\{x\in D'\right\}}}{\text{Leb}\left(D'\right)}, \qquad g_1\left(x\right) = \frac{1_{\left\{x\in D\right\}}}{\text{Leb}\left(D\right)},
\end{equation*}
where $\text{Leb}$ denotes the Lebesgue measure. Applying (\ref{eq:adoubleint}) to these choices of $g_0$ and $g_1$, we obtain
\begin{eqnarray}
&\,&\frac{1}{\text{Leb}\left(D\right)^2} \iint 1_{\left\{\left(x^1,x^2\right)\in A_{11}\right\}}1_{\left\{x^1\in D\right\}}1_{\left\{x^2\in D\right\}}dx^1dx^2\nonumber\\
&=& \frac{1}{\text{Leb}\left(D\right)}\frac{1}{\text{Leb}\left(D'\right)}\iint 1_{\left\{\left(x^1,x^2\right)\in A_{01}\right\}}1_{\left\{x^1\in D'\right\}}1_{\left\{x^2\in D\right\}}dx^1dx^2.\label{eq:intswithd}
\end{eqnarray}
Since $D'\times D \subseteq A_{01}$, (\ref{eq:intswithd}) equals $1$. Consequently, the set $A_{11}\cap \left(D\times D\right)$ has Lebesgue measure $\text{Leb}\left(D\right)^2$. This is only possible if $D\times D \subseteq A_{11}$. Since $x \in D$, it follows that $\left(x,x\right) \in D\times D$.

\noindent\textit{Step 2:} We show that $\mathcal{X}\times J = A_{01}$.

We proceed by contradiction: fix $x\in J$ and suppose that there exists some $x'\in\mathcal{X}$ such that $\left(x',x\right)\notin A_{01}$. Since $A_{01}$ can be assumed to be open, there exist open intervals $F',F$ such that $\left(x',x\right) \in F'\times F$ and $A_{01}\cap \left(F'\times F\right) = \emptyset$. Applying Step 1, we obtain $D^*$ such that $D^*\times D^*\subseteq A_{11}$ and $\left(x,x\right)\in D^*\times D^*$. Without loss of generality, we can assume $D^*\subseteq F$.

Let $g_1\left(x\right) = \frac{1}{\text{Leb}\left(D^*\right)}1_{\left\{x\in D^*\right\}}$. Applying (\ref{eq:adoubleint}) to this $g_1$ and any choice of $g_0$, we obtain
\begin{eqnarray*}
&\,&\frac{1}{\text{Leb}\left(D^*\right)^2} \iint 1_{\left\{\left(x^1,x^2\right)\in A_{11}\right\}}1_{\left\{x^1\in D^*, x^2 \in D^*\right\}}dx^1dx^2\\
&=& \frac{1}{\text{Leb}\left(D^*\right)}\iint 1_{\left\{\left(x^1,x^2\right)\in A_{01}\right\}}g_0\left(x^1\right)1_{\left\{x^2\in D^*\right\}}dx^1dx^2.
\end{eqnarray*}
Since $D^*\times D^*\subseteq A_{11}$, the left-hand side equals $1$. Thus,
\begin{equation*}
\int g_0\left(x^1\right) \left(\frac{1}{\text{Leb}\left(D^*\right)}\int 1_{\left\{\left(x^1,x^2\right)\in A_{01}\right\}} 1_{\left\{x^2\in D^*\right\}}dx^2\right)dx^1=1,
\end{equation*}
for any density $g_0$. It follows that
\begin{equation}\label{eq:intsforallx}
\frac{1}{\text{Leb}\left(D^*\right)}\int 1_{\left\{\left(x^1,x^2\right)\in A_{01}\right\}} 1_{\left\{x^2\in D^*\right\}}dx^2=1, \qquad \forall x^1\in\mathcal{X}.
\end{equation}
Specifically, (\ref{eq:intsforallx}) also holds if we take $x^1 = x'$. However, from before, we have $\left(x',x^2\right) \in F'\times D^*$, and $A_{01}\cap \left(F'\times D^*\right) = \emptyset$, which makes it impossible to satisfy (\ref{eq:intsforallx}). Therefore, $\mathcal{X}\times J \subseteq A_{01}$. By the definition of $J$ in (\ref{eq:Jdef}), we also have $A_{01}\subseteq \mathcal{X}\times J$, which yields the desired result.

\textit{Step 3:} We show that $J\times J = A_{11}$.

Since $A_{01} = \mathcal{X}\times J$, (\ref{eq:adoubleint}) implies that
\begin{equation*}
\iint 1_{\left\{\left(x^1,x^2\right)\in A_{11}\right\}}g_1\left(x^1\right)g_1\left(x^2\right)dx^1 dx^2 = \int 1_{\left\{x^2 \in J\right\}}g_1\left(x^2\right)dx^2
\end{equation*}
for all $g_1$. Taking $g_1\left(x\right) = \frac{1}{\text{Leb}\left(J\right)}1_{\left\{x\in J\right\}}$ implies $J\times J \subseteq A_{11}$. Applying Step 1, we also have $A_{11}\subseteq J\times J$.

\textit{Step 4:} We show that $J = \mathcal{X}$.

Since $A_{01} =\mathcal{X}\times J$ and $A_{11} = J\times J$, (\ref{eq:adoubleint}) implies that
\begin{equation*}
\left(\int_J g_1\left(x\right)dx\right)^2 = \int_J g_1\left(x\right)dx,
\end{equation*}
which means that $\int_J g_1\left(x\right)dx \in \left\{0,1\right\}$ for any choice of $g_1$. This is only possible if $J =\mathcal{X}$.

\subsection{Proof of Theorem \ref{thm:minimax}}

Fix an arbitrary $\beta_0\neq 0$ and define $\beta_1 = \beta_0 + n^{-\frac{1}{2}}v$, where $v\in\mathds{R}^p$ is chosen to satisfy $\beta^\top_0 v = 0$ and $\|v\| < 1$.

For notational convenience, let $\mathcal{H}$ be the support of the historical data $\bm{H}_n$, and similarly let $\mathcal{W}$ be the support of the new data $\bm{W} = \left(\bm{X},\bm{Z}\right)$. Analogously, let $p^{H}_{\beta}$ and $p^{W}_{\beta}$ be the probability measures induced by $\bm{H}_n$ and $\bm{W}$ on $\mathcal{H}$ and $\mathcal{W}$, respectively, under the parameter value $\beta$.

Define the set
\begin{equation}\label{eq:defofD}
D = \left\{w\in\mathcal{W}\,:\, I_{\beta_0}\left(w\right)=0, \,I_{\beta_1}\left(w\right)=1\right\},
\end{equation}
and the constant
\begin{equation}\label{eq:defofdelta}
\delta = P_{\beta_0}\left(\bm{W} \in D\right) = \int_{D} p^W_{\beta_0}\left(dw\right).
\end{equation}
Observe that, for $\beta\in\left\{\beta_0,\beta_1\right\}$,
\begin{eqnarray*}
P_{\beta}\left(\phi\left(\bm{H}_n,\bm{W}\right)\neq I_\beta\left(\bm{W}\right)\right) &=& \int_{\mathcal{H}}\int_{\mathcal{W}}1_{\left\{\phi\left(h,w\right)\neq I_{\beta}\left(w\right)\right\}}p^H_{\beta}\left(dh\right) p^W_{\beta}\left(dw\right)\\
&\geq& \int_{\mathcal{H}}\int_{D}1_{\left\{\phi\left(h,w\right)\neq I_{\beta}\left(w\right)\right\}}p^H_{\beta}\left(dh\right) p^W_{\beta}\left(dw\right).
\end{eqnarray*}
Therefore,
\begin{eqnarray}
&\,& P_{\beta_0}\left(\phi\left(\bm{H}_n,\bm{W}\right)\neq I_{\beta_0}\left(\bm{W}\right)\right) + P_{\beta_1}\left(\phi\left(\bm{H}_n,\bm{W}\right)\neq I_{\beta_1}\left(\bm{W}\right)\right)\nonumber\\
&\geq & \int_{\mathcal{H}}\int_D 1_{\left\{\phi\left(h,w\right)\neq I_{\beta_0}\left(w\right)\right\}}p^H_{\beta_0}\left(dh\right)p^W_{\beta_0}\left(dw\right) + 1_{\left\{\phi\left(h,w\right)\neq I_{\beta_1}\left(w\right)\right\}}p^H_{\beta_1}\left(dh\right)p^W_{\beta_1}\left(dw\right)\nonumber\\
&=& \int_{\mathcal{H}}\int_D \left(1_{\left\{\phi\left(h,w\right)\neq I_{\beta_0}\left(w\right)\right\}}p^H_{\beta_0}\left(dh\right) + 1_{\left\{\phi\left(h,w\right)\neq I_{\beta_1}\left(w\right)\right\}}p^H_{\beta_1}\left(dh\right)\right)p^W_{\beta_0}\left(dw\right)\label{eq:samedist}\\
&=& \int_{\mathcal{H}}\int_D \left(1_{\left\{\phi\left(h,w\right) = 1\right\}}p^H_{\beta_0}\left(dh\right) + 1_{\left\{\phi\left(h,w\right) = 0\right\}}p^H_{\beta_1}\left(dh\right)\right)p^W_{\beta_0}\left(dw\right)\label{eq:switchto01}\\
&=& \int_{\mathcal{H}}\int_D \left(p^H_{\beta_1}\left(dh\right) + 1_{\left\{\phi\left(h,w\right)=1\right\}}\left(p^H_{\beta_0}\left(dh\right)-p^H_{\beta_1}\left(dh\right)\right)\right)p^W_{\beta_0}\left(dw\right)\nonumber\\
&=& \int_{\mathcal{H}}\int_D p^H_{\beta_1}\left(dh\right)p^W_{\beta_0}\left(dw\right) + \int_{\mathcal{H}}\int_D 1_{\left\{\phi\left(h,w\right)=1\right\}}\left(p^H_{\beta_0}\left(dh\right)-p^H_{\beta_1}\left(dh\right)\right)p^W_{\beta_0}\left(dw\right)\nonumber\\
&=& \delta + \int_{\mathcal{H}}\int_D 1_{\left\{\phi\left(h,w\right)=1\right\}}\left(p^H_{\beta_0}\left(dh\right)-p^H_{\beta_1}\left(dh\right)\right)p^W_{\beta_0}\left(dw\right)\label{eq:delta1}\\
&\geq& \delta - \int_{\mathcal{H}}\int_D\left|p^H_{\beta_0}\left(dh\right)-p^H_{\beta_1}\left(dh\right)\right|p^W_{\beta_0}\left(dw\right)\nonumber\\
&=& \delta - \left(\int_{\mathcal{H}}\left|p^H_{\beta_0}\left(dh\right)-p^H_{\beta_1}\left(dh\right)\right|\right)\left(\int_{D} p^W_{\beta_0}\left(dw\right)\right)\nonumber\\
&=& \delta\left(1 - \int_{\mathcal{H}}\left|p^H_{\beta_0}\left(dh\right)-p^H_{\beta_1}\left(dh\right)\right|\right).\label{eq:delta2}
\end{eqnarray}
Above, (\ref{eq:samedist}) follows from the fact that $p^W_{\beta_0} = p^W_{\beta_1}$, because the distribution of $\left(X,Z\right)$ does not depend on the regression coefficients. Equation (\ref{eq:switchto01}) follows from the definition of $D$ in (\ref{eq:defofD}), while both (\ref{eq:delta1}) and (\ref{eq:delta2}) follow from the definition of $\delta$ in (\ref{eq:defofdelta}).

By Pinsker's inequality (see, e.g., Lemma 2.5 of \citealp{Ts09}), we have
\begin{equation}\label{eq:pinsker}
\int_{\mathcal{H}}\left|p^H_{\beta_0}\left(dh\right)-p^H_{\beta_1}\left(dh\right)\right| \leq \sqrt{2 D^{KL}\left(p^H_{\beta_0},p^H_{\beta_1}\right)},
\end{equation}
where $D^{KL}$ denotes the Kullback-Leibler (KL) divergence between two probability distributions. The inequality can be applied because the left-hand side of (\ref{eq:pinsker}) is a scalar multiple of the so-called total variation distance between $p^H_{\beta_0}$ and $p^H_{\beta_1}$.

We now calculate the KL divergence. For any $\beta \in \left\{\beta_0,\beta_1\right\}$ and $h=\left(x,z,y\right)$, we can decompose $p^H_{\beta}\left(h\right) = p^{X,Y}_{\beta}\left(x,y\right)p^Z\left(z\right)$, because we have assumed $X\perp Z$ in the construction of our setting. Note also that the distribution of $Z$ is not affected by $\beta$. Consequently, by the properties of the KL divergence, we have
\begin{equation*}
D^{KL}\left(p^H_{\beta_0},p^H_{\beta_1}\right) = D^{KL}\left(p^{X,Y}_{\beta_0},p^{X,Y}_{\beta_1}\right).
\end{equation*}
Due to the way we have constructed our setting, $\left(X,Y\right) \sim \mathcal{N}\left(0,\Sigma_{\beta}\right)$ with
\begin{equation*}
\Sigma_{\beta} = \left(
\begin{array}{c c}
\bm{I}_p & \beta\\
\beta^\top & \|\beta\|^2_2 + \sigma^2
\end{array}
\right).
\end{equation*}
Since $p^{X,Y}_{\beta}$ is the product measure ($n$-fold) of the $\mathcal{N}\left(0,\Sigma_{\beta}\right)$ distribution, we easily obtain
\begin{equation*}
D^{KL}\left(p^{X,Y}_{\beta_0},p^{X,Y}_{\beta_1}\right) = \frac{n}{2}\left(\tr\left(\Sigma^{-1}_{\beta_0}\Sigma_{\beta_1}\right) - \left(p+1\right)+ \log\left(\frac{\det \Sigma_{\beta_0}}{\det \Sigma_{\beta_1}}\right)\right).
\end{equation*}
Because the vector $v$ used to define $\beta_1$ was chosen to satisfy $\beta^\top_0 v = 0$, we then obtain
\begin{eqnarray*}
\det \Sigma_{\beta_0} &=& \det \Sigma_{\beta_1} = \sigma^2,\\
\tr\left(\Sigma^{-1}_{\beta_0}\Sigma_{\beta_1}\right) &=& \frac{1}{n}\|v\|^2_2,
\end{eqnarray*}
where $D^{KL}\left(p^{X,Y}_{\beta_0},p^{X,Y}_{\beta_1}\right) = \frac{1}{2}\|v\|^2_2$. Combining these results with (\ref{eq:delta2}) yields
\begin{equation}\label{eq:hineq}
P_{\beta_0}\left(\phi\left(\bm{H}_n,\bm{W}\right)\neq I_{\beta_0}\left(\bm{W}\right)\right) + P_{\beta_1}\left(\phi\left(\bm{H}_n,\bm{W}\right)\neq I_{\beta_1}\left(\bm{W}\right)\right) \geq \delta\left(1 - \|v\|_2\right).
\end{equation}

This bound can be connected to $n$ by further analyzing $\delta$. Let $g$ denote the standard normal pdf, and let $c>0$ be a constant. Then,
\begin{eqnarray}
\delta &=& P_{\beta_0}\left(\beta^\top_0 \left(X^1-X^2\right) < 0, \; \beta^\top_1 \left(X^1-X^2\right) \geq 0\right)\nonumber\\
&=& P_{\beta_0}\left(n^{-\frac{1}{2}}v^\top\left(X^1-X^2\right) \leq \beta^\top_0 \left(X^1-X^2\right) < 0\right)\nonumber\\
&\geq& P_{\beta_0}\left(n^{-\frac{1}{2}}v^\top\left(X^1-X^2\right) \leq \beta^\top_0 \left(X^1-X^2\right) < 0, \quad -c\leq v^\top \left(X^1-X^2\right) \leq \frac{c}{2}\right)\nonumber\\
&\geq& P_{\beta_0}\left(n^{-\frac{1}{2}}\frac{c}{2} \leq \beta^\top_0 \left(X^1-X^2\right) < 0, \quad -c\leq v^\top \left(X^1-X^2\right) \leq \frac{c}{2}\right)\nonumber\\
&=& P_{\beta_0}\left(n^{-\frac{1}{2}}\frac{c}{2} \leq \beta^\top_0 \left(X^1-X^2\right) < 0\right)P_{\beta_0}\left(-c\leq v^\top \left(X^1-X^2\right) \leq \frac{c}{2}\right),\label{eq:cindep}
\end{eqnarray}
where (\ref{eq:cindep}) is due to the fact that $\beta^\top_0\left(X^1-X^2\right)\perp v^\top\left(X^1-X^2\right)$ because $\bm{X}$ is jointly Gaussian and $\beta^\top_0 v = 0$. The second probability in (\ref{eq:cindep}) has no dependence on $n$. For the first probability, we can use the bound
\begin{eqnarray}
P_{\beta_0}\left(n^{-\frac{1}{2}}\frac{c}{2} \leq \beta^\top_0 \left(X^1-X^2\right) < 0\right) &=& P_{\beta_0}\left( \frac{n^{-\frac{1}{2}}c}{2\sqrt{2} \|\beta_0\|_2} \leq \frac{\beta^\top_0 \left(X^1-X^2\right)}{\sqrt{2} \|\beta_0\|_2} < 0\right)\nonumber\\
&\geq& \left(\frac{n^{-\frac{1}{2}}c}{2\sqrt{2} \|\beta_0\|_2}\right)g\left(\frac{n^{-\frac{1}{2}}c}{2\sqrt{2} \|\beta_0\|_2}\right),\label{eq:stdpdf}
\end{eqnarray}
where $g$ is the standard normal pdf. The final inequality (\ref{eq:stdpdf}) is obtained by first scaling $\beta^\top_0\left(X^1-X^2\right)$ to make it standard normal, then applying the bound $\int^a_0 g\left(t\right)dt \geq a\cdot g\left(a\right)$. Combining (\ref{eq:cindep}) with (\ref{eq:stdpdf}), we conclude that there exists $\kappa$, whose value depends on $c,\beta_0,v$ but not on $n$, satisfying $\delta \geq n^{-\frac{1}{2}}\kappa$. Together with (\ref{eq:hineq}), this yields
\begin{equation*}
P_{\beta_0}\left(\phi\left(\bm{H}_n,\bm{W}\right)\neq I_{\beta_0}\left(\bm{W}\right)\right) + P_{\beta_1}\left(\phi\left(\bm{H}_n,\bm{W}\right)\right) \geq n^{-\frac{1}{2}}\kappa\left(1-\|v\|_2\right).
\end{equation*}
Since
\begin{equation*}
\max_{\beta\in\left\{\beta_0,\beta_1\right\}} P_{\beta}\left(\phi\left(\bm{H}_n,\bm{W}\right)\neq I_{\beta}\left(\bm{W}\right)\right) \geq \frac{1}{2}\left( P_{\beta_0}\left(\phi\left(\bm{H}_n,\bm{W}\right)\neq I_{\beta_0}\left(\bm{W}\right)\right) + P_{\beta_1}\left(\phi\left(\bm{H}_n,\bm{W}\right)\right)\right),
\end{equation*}
this completes the proof.

\subsection{Proof of Theorem \ref{thm:regconc}}\label{sec:regtail}

To do this, we consider the event
\begin{equation*}
E = \left\{\lambda_{\min}\left(\frac{1}{n}\sum^n_{m=1} X_mX^\top_m\right) \geq \frac{1}{2}\kappa_2\right\},
\end{equation*}
where $\lambda_{\min}\left(\cdot\right)$ denotes the smallest eigenvalue of a matrix. We will then argue that
\begin{eqnarray}
P\left(E^c\right) &\leq& C_1 e^{-C_2n},\label{eq:eigbound}\\
P\left(E\cap\left\{\sqrt{n}\|\hat{\beta}-\beta\|_2 > s\right\}\right) &\leq& C_1 e^{-C_2 s^2},\label{eq:regbound}
\end{eqnarray}
which together imply (\ref{eq:regconc}).

We first focus on (\ref{eq:eigbound}). From Assumption \ref{a1}(iii), $\left|X_{m,j}\right|\leq\kappa_4$ for all $m=1,...,n$, $j=1,...,p$. Consequently, the sub-Gaussian norm of the product $X_{m,j}X_{m,j'}$ satisfies $\|X_{m,j}X_{m,j'}\|_{SG} \leq C$ where $C$ is a constant whose value depends only on $\kappa_4$. By a Hoeffding-type inequality (e.g., Prop. 5.10 of \citealp{Ve12}), we have
\begin{equation*}
P\left(\left|\sum^n_{m=1} \left(X_{m,j}X_{m,j'} - \mathbb{E}\left(X_{m,j}X_{m,j'}\right)\right)\right|>s\right) \leq C'e^{-C'' s^2 n^{-1}},
\end{equation*}
for any $s > 0$ and any $j,j'$, with the constants $C',C''$ depending only on $C$. Recall that, for any matrix $A$, the spectral norm $\|A\|_2 = \sqrt{\lambda_{\max}\left(A^\top A\right)}$ is bounded above by the Frobenius norm, i.e., $\|A\|_2 \leq \|A\|_F$. Applying a union bound, we derive
\begin{eqnarray*}
P\left(\|\frac{1}{n}\sum^n_{m=1} X_mX^\top_m - \mathbb{E}\left(XX^\top\right)\|_2 > s\right) &\leq& P\left(\|\frac{1}{n}\sum^n_{m=1} X_mX^\top_m - \mathbb{E}\left(XX^\top\right)\|^2_F > s^2\right)\\
&=& P\left(\sum_{j,j'} \left|\frac{1}{n}\sum^n_{m=1} X_{m,j}X_{m,j'}-\mathbb{E}\left(X_{m,j}X_{m,j'}\right)\right|^2 > s^2\right)\\
&\leq& \sum_{j,j'} P\left(\left|\frac{1}{n}\sum^n_{m=1} X_{m,j}X_{m,j'}-\mathbb{E}\left(X_{m,j}X_{m,j'}\right)\right|^2 > p^{-2}s^2\right)\\
&=& \sum_{j,j'} P\left(\left|\sum^n_{m=1} X_{m,j}X_{m,j'}-\mathbb{E}\left(X_{m,j}X_{m,j'}\right)\right| > n p^{-1}s\right)\\
&\leq& p^2 C' e^{-C'' np^{-2}s^2}.
\end{eqnarray*}
Taking $s = \frac{1}{3}\kappa_2$, we have
\begin{equation*}
P\left(\|\frac{1}{n}\sum^n_{m=1} X_mX^\top_m - \mathbb{E}\left(XX^\top\right)\|_2 > \frac{1}{3}\kappa_2\right) \leq p^2 C' e^{-\frac{C''}{9}\kappa^2_2 np^{-2}}.
\end{equation*}
Recall the elementary inequality $\lambda_{\min}\left(A+B\right) \geq \lambda_{\min}\left(A\right) - \|B\|_2$ for any two symmetric matrices $A,B$. Therefore,
\begin{eqnarray}
P\left(\lambda_{\min}\left(\sum^n_{m=1} X_mX^\top_m\right)\leq \frac{1}{2}\kappa_2\right) &\leq& P\left(\lambda_{\min}\left(\mathbb{E}\left(XX^\top\right)\right) - \|\frac{1}{n}\sum^n_{m=1}X_mX^\top_m -\mathbb{E}\left(XX^\top\right)\|_2 \leq \frac{1}{2}\kappa_2\right)\nonumber\\
&\leq& P\left(\kappa_2 - \|\frac{1}{n}\sum^n_{m=1}X_mX^\top_m -\mathbb{E}\left(XX^\top\right)\|_2 \leq \frac{1}{2}\kappa_2\right)\label{eq:eigbound1}\\
&=& P\left(\|\frac{1}{n}\sum^n_{m=1}X_mX^\top_m -\mathbb{E}\left(XX^\top\right)\|_2 \geq \frac{1}{2}\kappa_2\right)\nonumber\\
&\leq& p^2 C'e^{-\frac{C''}{9} \kappa^2_2 n p^{-2}},\nonumber
\end{eqnarray}
where (\ref{eq:eigbound1}) follows by Assumption \ref{a1}(ii). This completes the proof of (\ref{eq:eigbound}).

We now turn to (\ref{eq:regbound}). Since $X$ is bounded and $\varepsilon$ is sub-Gaussian, $X_j\varepsilon$ is sub-Gaussian for any $j$, with $\mathbb{E}\left(X_j\varepsilon\right) = 0$. By a Hoeffding-type inequality (e.g., Prop. 5.10 of \citealp{Ve12}), we have
\begin{equation*}
P\left(\left|\sum^n_{m=1}X_{m,j}\varepsilon_m\right| > s\sqrt{\frac{n}{p}}\right) \leq c e^{-c' \frac{s^2}{p}},
\end{equation*}
for any $s > 0$ and any $j$, with $c,c'$ being constants whose values depend only on $\kappa_4,\kappa_5$. Then, for any $s>0$,
\begin{eqnarray*}
P\left(\sqrt{n}\|\frac{1}{n}\sum^n_{m=1} X_m\varepsilon_m\|_2 > s\right) &=& P\left(n\|\frac{1}{n}\sum^n_{m=1} X_m\varepsilon_m\|^2_2 > s^2\right)\\
&=& P\left(\sum^p_{j=1} \frac{1}{n}\left|\sum^n_{m=1} X_{m,j}\varepsilon_m\right|^2 > s^2\right)\\
&\leq& \sum^p_{j=1} P\left(\frac{1}{n}\left|\sum^n_{m=1} X_{m,j}\varepsilon_m\right|^2 > \frac{s^2}{p}\right)\\
&\leq& \sum^p_{j=1} P\left(\left|\sum^n_{m=1} X_{m,j}\varepsilon_m\right| > s\sqrt{\frac{n}{p}}\right)\\
&\leq& c p e^{-c' \frac{s^2}{p}}.
\end{eqnarray*}
Finally, we write
\begin{eqnarray*}
P\left(E\cap \left\{\left\{\sqrt{n}\|\hat{\beta}-\beta\|_2 > s\right\}\right\}\right) &\leq& P\left(\frac{1}{2}\sqrt{n}\|\frac{1}{n}\sum^n_{m=1}X_m\varepsilon_m\|_2 \kappa_2 > s\right)\\
&=& P\left(\sqrt{n}\|\frac{1}{n}\sum^n_{m=1}X_m\varepsilon_m\|_2 > 2\frac{s}{\kappa_2}\right)\\
&\leq& cp e^{-4c' \frac{s^2}{p\kappa^2_2}},
\end{eqnarray*}
completing the proof of (\ref{eq:regbound}).

\subsection{Proof of Theorem \ref{thm:cdfconc}}\label{sec:cdftail}

Without loss of generality, we can let $z = 1$, as the other case $z=0$ can be obtained by an analogous argument. We can then simplify the notation by letting $F_\theta = F_{1,\theta}$ and $\hat{F}_{\theta} = \hat{F}_{1,\theta}$.

Recall the notation $\rho = P\left(Z=1\right)$ and define $\Delta = \frac{1}{n}\sum^n_{m=1} Z_m - \rho$. We observe that
\begin{eqnarray}
\hat{F}_\theta\left(r\right) - F_\theta\left(r\right) &=& \frac{\frac{1}{n}\sum^n_{m=1} Z_m 1_{\left\{\theta^\top X_m\leq r\right\}}}{\rho+\Delta} - \frac{\mathbb{E}\left(Z 1_{\left\{\theta^\top X\leq r\right\}}\right)}{\rho}\nonumber\\
&=& -\frac{\Delta}{\rho\left(\rho+\Delta\right)}\frac{1}{n}\sum^n_{m=1} Z_m 1_{\left\{\theta^\top X_m\leq r\right\}} + \frac{1}{\rho}\left(\frac{1}{n}\sum^n_{m=1} Z_m 1_{\left\{\theta^\top X_m\leq r\right\}} - \mathbb{E}\left(Z 1_{\left\{\theta^\top X\leq r\right\}}\right)\right).\label{eq:Fbound1}
\end{eqnarray}
We will bound both terms in (\ref{eq:Fbound1}). First, since $Z$ is bounded, we can apply Hoeffding's inequality (see, e.g., Thm. 2.8 of \citealp{BoLuMa13}) and obtain
\begin{equation}\label{eq:hoeffding}
P\left(\left|\Delta\right|\geq a n^{-\frac{1}{2}}\right) \leq e^{-\frac{2a^2}{C}}
\end{equation}
for any $a>0$, with $C$ being a constant whose value depends only on $\kappa_1$. Using Assumption \ref{a1}(i), we then derive
\begin{eqnarray}
&\,& P\left(\sup_r \left|\frac{\Delta}{\rho\left(\rho+\Delta\right)}\frac{1}{n}\sum^n_{m=1} Z_m 1_{\left\{\theta^\top X_m\leq r\right\}}\right| > an^{-\frac{1}{2}}, \quad \left|\Delta\right| \leq \frac{\rho}{2}\right)\nonumber\\
&\leq& P\left(\frac{\Delta}{\rho\left(\rho+\Delta\right)} > an^{-\frac{1}{2}}, \quad \left|\Delta\right|\leq \frac{\rho}{2}\right)\label{eq:Fbound2}\\
&\leq& P\left(\frac{2\Delta}{\rho^2} > an^{-\frac{1}{2}}\right)\nonumber\\
&\leq& e^{-\frac{a^2 \rho^4}{2C}}\label{eq:Fbound3},
\end{eqnarray}
where (\ref{eq:Fbound2}) is due to the fact that $\sup_r \left|\sum^n_{m=1} Z_m 1_{\left\{\theta^\top X_m\leq r\right\}}\right|\in\left\{0,1\right\}$, and (\ref{eq:Fbound3}) follows by (\ref{eq:hoeffding}). Now,
\begin{eqnarray}
&\,& P\left(\sup_r \left|\frac{\Delta}{\rho\left(\rho+\Delta\right)}\frac{1}{n}\sum^n_{m=1} Z_m 1_{\left\{\theta^\top X_m\leq r\right\}}\right| > an^{-\frac{1}{2}}\right)\nonumber\\
&\leq& P\left(\sup_r \left|\frac{\Delta}{\rho\left(\rho+\Delta\right)}\frac{1}{n}\sum^n_{m=1} Z_m 1_{\left\{\theta^\top X_m\leq r\right\}}\right| > an^{-\frac{1}{2}}, \quad \left|\Delta\right| \leq \frac{\rho}{2}\right) + P\left(\left|\Delta\right| > \frac{\rho}{2}\right)\nonumber\\
&\leq& e^{-\frac{a^2 \rho^4}{2C}} + e^{-\frac{n\rho^2}{2C}}\label{eq:Fbound4}\\
&\leq& e^{-\frac{a^2 \kappa^4_1}{2C}}+e^{-\frac{n\kappa^2_1}{2C}},\label{eq:Fbound5}
\end{eqnarray}
where (\ref{eq:Fbound4}) combines (\ref{eq:Fbound3}) with another application of (\ref{eq:hoeffding}), and (\ref{eq:Fbound5}) follows from Assumption \ref{a1}.

Now, we return to (\ref{eq:Fbound1}) and bound the second term. Let $\mathcal{F}$ be the class of functions of the form $X\mapsto r - \theta^\top X$ for some $\left(\theta,r\right)\in\mathds{R}^p\times\mathds{R}$. This is a finite-dimensional class with dimension $p+1$. By Lemma 2.6.15 of \cite{VaWe96}, $\mathcal{F}$ is a Vapnik-Chervonenkis class (VC-class). By Lemma 2.6.18(iii) of \cite{VaWe96}, the class of functions of the form $X\mapsto 1_{\left\{r-\theta^\top X > 0\right\}}$ is also a VC-class. By part (vi) of the same lemma, the class $\mathcal{G}$ of functions of the form $\left(X,Z\right)\mapsto Z1_{\left\{r-\theta^\top X > 0\right\}}$ is also a VC-class.

Let $V$ denote the Vapnik-Chervonenkis dimension (VC-dimension) of $\mathcal{G}$. A consequence of the preceding arguments is that $V$ is bounded above by a constant whose value depends only on $p$. Additionally, the constant function always taking the value $1$ is an envelope function for $\mathcal{G}$. Applying Theorems 2.6.7 and 2.14.9 of \cite{VaWe96}, we obtain the inequality
\begin{equation*}
P\left(\sqrt{n}\sup_{\left(\theta,r\right)} \left|\frac{1}{n}\sum^n_{m=1} Z_m 1_{\left\{\theta^\top X_m\leq r\right\}} - \mathbb{E}\left(Z 1_{\left\{\theta^\top X\leq r\right\}}\right)\right| > b \right) \leq C' b^V e^{-2b^2}
\end{equation*}
for any $b > 0$, with $C'$ being a constant whose value depends on $V$. Then,
\begin{equation}\label{eq:Fbound6}
P\left(\sqrt{n}\sup_{\left(\theta,r\right)} \frac{1}{\rho}\left|\frac{1}{n}\sum^n_{m=1} Z_m 1_{\left\{\theta^\top X_m\leq r\right\}} - \mathbb{E}\left(Z 1_{\left\{\theta^\top X\leq r\right\}}\right)\right| > b \right) \leq C' \rho^V b^V e^{-2\rho^2 b^2}.
\end{equation}
We now combine (\ref{eq:Fbound1}) with (\ref{eq:Fbound5}) and (\ref{eq:Fbound6}). For any $a,b>0$, we have
\begin{equation}\label{eq:Fbound7}
P\left(\sup_{\left(\theta,r\right)}\left|\hat{F}_\theta\left(r\right) - F_\theta\left(r\right)\right| > an^{-\frac{1}{2}} + bn^{-\frac{1}{2}}\right) \leq e^{-\frac{a^2 \kappa^4_1}{2C}}+e^{-\frac{n\kappa^2_1}{2C}} + C' \rho^V b^V e^{-2\rho^2 b^2}.
\end{equation}
Let us take $b=t$ with $t\in\left(0,\frac{\sqrt{n}}{2\kappa_1}\right)$, and $a = t + \frac{1}{2\kappa_1}$. Then, (\ref{eq:Fbound7}) becomes
\begin{eqnarray}
&\,& P\left(\sup_{\left(\theta,r\right)}\left|\hat{F}_\theta\left(r\right) - F_\theta\left(r\right)\right| > \left(t+ \frac{1}{2\kappa_1}\right)n^{-\frac{1}{2}} + n^{-\frac{1}{2}} t\right)\nonumber\\
&\leq& e^{-\frac{\left(t+\frac{1}{2\kappa_1}\right)^2 \kappa^4_1}{2C}}+e^{-\frac{n\kappa^2_1}{2C}} + C' \rho^V b^V e^{-2\rho^2 t^2}\nonumber\\
&\leq& e^{-\frac{\left(t+\frac{1}{2\kappa_1}\right)^2 \kappa^4_1}{2C}}+e^{-\frac{\left(1+\frac{2}{\kappa_1}\right)^2\kappa^4_1}{2C}} + C' \rho^V b^V e^{-2\rho^2 t^2}\label{eq:Fbound8}\\
&\leq& 2e^{-\frac{t^2 \kappa^4_1}{2C}} + C' \rho^V b^V e^{-2\rho^2 t^2},\nonumber
\end{eqnarray}
where (\ref{eq:Fbound8}) is due to the fact that $t \leq \frac{\sqrt{n}}{2\kappa_1}$ and
\begin{equation*}
\left(t+\frac{1}{2\kappa_1}\right)^2\kappa^4_1 \leq 2t^2 \kappa^4_1 + \frac{2\kappa^4_1}{\left(2\kappa_1\right)^2} \leq n\kappa^4_2.
\end{equation*}
The desired result (\ref{eq:cdfconc}) then follows by choosing sufficiently large constants.

\subsection{Proof of Theorem \ref{thm:Tconc}}

We first state a technical result that will be useful further down. The proof is given in a separate section of the Appendix.

\begin{lem}\label{lem:tech1}
For $z\in\left\{0,1\right\}$ and any $h$,
\begin{equation*}
\sup_r \left| F_{z,\theta+h}\left(r\right)-F_{z,\theta}\left(r\right)\right| \leq \kappa_4\kappa_6\|h\|_2.
\end{equation*}
\end{lem}

Recalling (\ref{eq:T}) and (\ref{eq:That}), we fix $t$ and write
\begin{eqnarray}
\left|\hat{T}\left(t\right)-T\left(t\right)\right| &=& \left|\int \left(\hat{F}_{1}\left(u+t\right)\right)^{K_1}\left(\hat{F}_0\left(u\right)\right)^{K_0-1}d\hat{F}_0\left(u\right) - \int\left(F_{1,\beta}\left(u+t\right)\right)^{K_1}\left(F_{0,\beta}\left(u\right)\right)^{K_0-1}dF_{0,\beta}\left(u\right)\right|\nonumber\\
&\leq& J_1\left(t\right) + J_2\left(t\right),\label{eq:J1J2}
\end{eqnarray}
where
\begin{eqnarray*}
J_1\left(t\right) &=& \left|\int \left(\hat{F}_1\left(u+t\right)\right)^{K_1}\left(\hat{F}_0\left(u\right)\right)^{K_0-1}d\hat{F}_0\left(u\right) - \int\left(\hat{F}_1\left(u+t\right)\right)^{K_1}\left(\hat{F}_0\left(u\right)\right)^{K_0-1}dF_{0,\beta}\left(u\right)\right|,\\
J_2\left(t\right) &=& \left|\int \left(\hat{F}_1\left(u+t\right)\right)^{K_1}\left(\hat{F}_0\left(u\right)\right)^{K_0-1}dF_{0,\beta}\left(u\right) - \int\left(F_{1,\beta}\left(u+t\right)\right)^{K_1}\left(F_{0,\beta}\left(u\right)\right)^{K_0-1}dF_{0,\beta}\left(u\right)\right|.
\end{eqnarray*}
We will bound $J_1\left(t\right)$ and $J_2\left(t\right)$ separately.

Let us first consider $J_1$. Let $\mathcal{F}$ denote the class of monotone functions mapping $\mathds{R}$ into $\left[0,1\right]$, and let
\begin{equation*}
W = \sqrt{n} \sup_{f\in\mathcal{F}}\left| \int f d\hat{F}_0 - \int f dF_{0,\beta}\right|.
\end{equation*}
Clearly, the mapping
\begin{equation*}
u\mapsto \left(\hat{F}_1\left(u+t\right)\right)^{K_1}\left(\hat{F}_0\left(u\right)\right)^{K_0-1}
\end{equation*}
belongs to $\mathcal{F}$ for any realization of the historical data. It follows that $\sqrt{n} \sup_t J_1\left(t\right) \leq W$.

For any $\delta>0$ and any probability measure $Q$, we let $N_{[]}\left(\delta,\mathcal{F},L_2\left(Q\right)\right)$ denote the bracketing number as in Definition 2.1.6 of \cite{VaWe96}. By Theorem 2.7.5 of \cite{VaWe96}, there exists a universal constant $D_1>0$ such that, for any $\delta > 0$ and any probability measure $Q$,
\begin{equation*}
\log N_{[]}\left(\delta,\mathcal{F},L_2\left(Q\right)\right) \leq D_1 \delta^{-1}.
\end{equation*}
Clearly, the constant function that equals $1$ everywhere is an envelope function for $\mathcal{F}$. By Theorem 2.14.2 of \cite{VaWe96}, there exists a universal constant $D_2>0$ such that
\begin{equation*}
\mathbb{E}\left(W\right) \leq D_2\int^1_0 \sqrt{1+\log N_{[]}\left(\delta,\mathcal{F},L_2\left(Q\right)\right)}d\delta \leq D_2 \int^1_0 \sqrt{1+\frac{D_1}{\delta}}d\delta.
\end{equation*}
Hence, $\mathbb{E}\left(W\right) \leq D_3$ for some universal constant $D_3>0$. By part 3 of Theorem 2.14.5 of \cite{VaWe96}, the sub-Gaussian norm of $W$ is bounded by a universal constant $D_4>0$. It follows that the sub-Gaussian norm of $\sqrt{n} \sup_t J_1\left(t\right)$ is also bounded by $D_4$. By the equivalence of sub-Gaussian norms and exponential bounds (e.g., Prop. 2.5.2 of \citealp{Ve18}), there exists a universal constant $D_5>0$ such that, for any $t>0$, the concentration inequality
\begin{equation}\label{eq:supofJ1}
P\left(\sqrt{n}\sup_t J_t\left(t\right) > t\right) \leq 2\exp\left(-D_5 t^2\right)
\end{equation}
holds. This establishes the desired bound on $J_1$.

Next, we consider $J_2$. The mapping $\left(a,b\right)\mapsto a^{K_1}b^{K_0-1}$ is Lipschitz on $\left[0,1\right]$. Let $D_6>0$ be a constant that bounds the Lipschitz norm of this mapping. Since $K_0,K_1\leq K$, we can choose $D_6$ in a way that depends on $K$ only. It follows that for any $u,t$,
\begin{eqnarray*}
&\,& \left|\left(\hat{F}_1\left(u+t\right)\right)^{K_1}\left(\hat{F}_0\left(u\right)\right)^{K_0-1}-\left(F_{1,\beta}\left(u+t\right)\right)^{K_1}\left(F_{0,\beta}\left(u\right)\right)^{K_0-1}\right|\\
&\leq& D_6\left(\|\hat{F}_1-F_{1,\beta}\|_{\infty} + \|\hat{F}_0-F_{0,\beta}\|_{\infty}\right),
\end{eqnarray*}
whence
\begin{equation}\label{eq:supofJ2}
\sup_t J_2\left(t\right) \leq D_6\left(\|\hat{F}_1-F_{1,\beta}\|_{\infty} + \|\hat{F}_0-F_{0,\beta}\|_{\infty}\right).
\end{equation}
For $z\in\left\{0,1\right\}$, we have
\begin{eqnarray}
\|\hat{F}_z-F_{z,\beta}\|_{\infty} &\leq & \|\hat{F}_{z,\hat{\beta}}-F_{z,\hat{\beta}}\|_{\infty} + \|F_{z,\hat{\beta}}-F_{z,\beta}\|_{\infty}\nonumber\\
&\leq& \sup_{\theta} \|\hat{F}_{z,\theta} - F_{z,\theta}\|_{\infty} + \|F_{z,\hat{\beta}}-F_{z,\beta}\|_{\infty}\nonumber\\
&\leq& \sup_{\theta} \|\hat{F}_{z,\theta} - F_{z,\theta}\|_{\infty} + \kappa_4\kappa_6\|\hat{\beta}-\beta\|_2\label{eq:applytechlem},
\end{eqnarray}
where (\ref{eq:applytechlem}) is due to Lemma \ref{lem:tech1}.

Applying Theorems \ref{thm:regconc} and \ref{thm:cdfconc}, we obtain, for $t\in\left(0,\frac{\sqrt{n}}{2\kappa_1}\right)$ and $z\in\left\{0,1\right\}$, the bound
\begin{eqnarray*}
&\,& P\left(\sqrt{n}\|\hat{F}_z-F_{z,\beta}\|_{\infty}>\left(1+\kappa_4\kappa_6\right)t + C_6\right)\\
&\leq& P\left(\sqrt{n}\sup_{\theta}\|\hat{F}_{z,\theta}-F_{z,\theta}\|_{\infty}> t + C_6\right) + P\left(\|\hat{\beta}-\beta\|_2 > t\right)\\
&\leq & C_7\exp\left(-C_8 t^2\right) + C_1\exp\left(-C_2 t^2\right) + C_1\exp\left(-C_2 n\right).
\end{eqnarray*}
By (\ref{eq:supofJ2}) and $\kappa_1\leq 1$, it follows that, for $t\in\left(0,\frac{\sqrt{n}}{2\kappa_1}\right)$, we have
\begin{equation*}
P\left(\sqrt{n}\sup_t J_2\left(t\right) > 2D_6\left(1+\kappa_4\kappa_6\right)t + 2D_6 C_6\right) \leq C_7\exp\left(-C_8 t^2\right) + C_1\exp\left(-C_2 t^2\right) + C_1\exp\left(-C_2 n\right).
\end{equation*}
Combining this bound with (\ref{eq:supofJ1}) and (\ref{eq:J1J2}) yields (\ref{eq:Tconc}) for $t\in\left(0,\frac{\sqrt{n}}{2}\right)$. Since the tail probability decreases as $t$ increases beyond $\frac{\sqrt{n}}{2}$, we can choose $C_9,C_{10},C_{11}>0$ to make (\ref{eq:Tconc}) hold for all $t>0$.

\subsection{Proof of Theorem \ref{thm:qconc}}

Recalling that $T\left(q\right)=0$ and $\hat{q}=\arg\min_t \left|\hat{T}\left(t\right)\right|$, we have
\begin{equation*}
\left|\hat{T}\left(\hat{q}\right)\right| \leq \left|\hat{T}\left(q\right)\right| = \left|\hat{T}\left(q\right)-T\left(q\right)\right| \leq \|\hat{T}-T\|_{\infty}.
\end{equation*}
Consequently,
\begin{equation*}
\left|T\left(\hat{q}\right)\right| \leq \left|\hat{T}\left(\hat{q}\right)\right| + \left|T\left(\hat{q}\right)-\hat{T}\left(\hat{q}\right)\right| \leq 2\|\hat{T}-T\|_{\infty}.
\end{equation*}
Since $T\left(q\right)=0$, we have
\begin{equation}\label{eq:Tbound}
\left|T\left(\hat{q}\right) - T\left(q\right)\right| \leq 2\|\hat{T}-T\|_{\infty}.
\end{equation}
Recall that $P\left(R^{(1)}-R^{(0)}\leq t\right) = K_0T\left(t\right) + \frac{K_0}{K}$. Then,
\begin{equation*}
T'\left(t\right) = K^{-1}_0\frac{d}{dt}P\left(R^{(1)}-R^{(0)}\leq t\right).
\end{equation*}
It follows that $T$ is an increasing function. By Assumption \ref{a1}(vi), $T'\left(t\right)$ is bounded below by $K^{-1}_0\kappa_8$ on the interval $q\pm \kappa_7$. It follows that
\begin{equation*}
T\left(q+\kappa_7\right) \geq T\left(q\right) + K^{-1}_0\kappa_7\kappa_8 \geq \kappa_7\kappa_8,
\end{equation*}
and, similarly, $T\left(q-\kappa_7\right)\leq -\kappa_7\kappa_8$. Therefore,
\begin{equation*}
P\left(\hat{q}-q\geq \kappa_7\right) = P\left(T\left(\hat{q}\right) \geq T\left(q+\kappa_7\right)\right) \leq P\left(T\left(\hat{q}\right)\geq \kappa_7\kappa_8\right).
\end{equation*}
Similarly, $P\left(\hat{q}-q\leq -\kappa_7\right) \leq P\left(T\left(\hat{q}\right)\leq -\kappa_7\kappa_8\right)$. Then, by (\ref{eq:Tbound}), we have
\begin{eqnarray}
P\left(\left|\hat{q}-q\right| \geq \kappa_7\right) &\leq& P\left(\left|T\left(\hat{q}\right)\right|\geq \kappa_7\kappa_8\right)\nonumber\\
&=& P\left(\left|T\left(\hat{q}\right)-T\left(q\right)\right|\geq \kappa_7\kappa_8\right)\nonumber\\
&\leq & P\left(2\|\hat{T}-T\|_{\infty} \geq \kappa_7\kappa_8\right).\label{eq:connecttoTconc}
\end{eqnarray}
Since $T'$ is bounded below by $K^{-1}_0\kappa_8$ on the interval $q\pm\kappa_7$, we have
\begin{equation*}
K^{-1}_0\kappa_8 \left|\hat{q}-q\right| \leq \left|\hat{T}\left(\hat{q}\right)-T\left(q\right)\right|
\end{equation*}
for $\left|\hat{q}-q\right|\leq\kappa_7$. Applying (\ref{eq:Tbound}), we obtain
\begin{equation*}
\left|\hat{q}-q\right| \leq \frac{2K_0}{\kappa_8}\|\hat{T}-T\|_{\infty} \leq \frac{2K}{\kappa_8}\|\hat{T}-T\|_{\infty}
\end{equation*}
for $\left|\hat{q}-q\right|\leq\kappa_7$.

Letting $C_9,C_{10},C_{11}$ be the constants obtained from Theorem \ref{thm:Tconc}, we have
\begin{eqnarray}
&\,& P\left(\sqrt{n}\left|\hat{q}-q\right|>t+C_{10}K\kappa^{-1}_8\right)\nonumber\\
&\leq & P\left(\left|\hat{q}-q\right|\geq\kappa_7\right) + P\left(2K\kappa^{-1}_8\sqrt{n}\|\hat{T}-T\|_{\infty} > t + C_{10}K\kappa^{-1}_8\right)\nonumber\\
&\leq & P\left(2\sqrt{n}\|\hat{T}-T\|_{\infty}\geq\sqrt{n}\kappa_7\kappa_8\right) + P\left(2\sqrt{n}\|\hat{T}-T\|_{\infty}>t\kappa_8 K^{-1}+C_{10}\right)\label{eq:qconc1}\\
&\leq& 2\exp\left(-C_{12}t^2\right) + 2\exp\left(-C_{12}n\right),\label{eq:qconc2}
\end{eqnarray}
where (\ref{eq:qconc1}) follows from (\ref{eq:connecttoTconc}), (\ref{eq:qconc2}) is due to Theorem \ref{thm:Tconc}, and $C_{12}$ is a constant depending on $C_9,C_{10},C_{11}$ as well as $\kappa_7,\kappa_8$.

Let $C_{13}$ be a small enough constant such that $C_{13}\leq\min\left\{C_{12},\frac{\kappa^2_8}{2C^2_{10}K^2}\right\}$ and also $C_{13}r^2 \leq C_{12}\left(r - C_{10}K\kappa^{-1}_8\right)^2$ for any $r \geq C_{10}K\kappa^{-1}_8$. Such a choice of $C_{13}$ exists and depends only on $C_{10},C_{12},\kappa_8,K$. The desired bound (\ref{eq:qconc}) then directly follows; to verify this, one can consider two cases $t < C_{10}K\kappa^{-1}_8$ and $t \geq C_{10}K\kappa^{-1}_8$ and apply the properties of $C_{13}$ together with (\ref{eq:qconc2}).

\subsection{Proof of Lemma \ref{lem:tech1}}

By definition,
\begin{equation*}
F_{z,\theta+h}\left(r\right) = \mathbb{E}\left(1_{\left\{\theta^\top X \leq r - h^\top X\right\}}\mid Z=z\right).
\end{equation*}
Observe that
\begin{equation*}
1_{\left\{\theta^\top X \leq r - h^\top X\right\}} - 1_{\left\{\theta^\top X \leq r\right\}} = \left\{
\begin{array}{c c}
-1 & r \geq \theta^\top X > r - h^\top X,\\
1 &  r < \theta^\top X \leq r - h^\top X,\\
0 & \mbox{otherwise.}
\end{array}
\right.
\end{equation*}
Therefore,
\begin{equation*}
\left|1_{\left\{\theta^\top X \leq r - h^\top X\right\}} - 1_{\left\{\theta^\top X \leq r\right\}}\right| \leq 1_{\left\{\left|\theta^\top X - r\right| \leq \left|h^\top X\right|\right\}},
\end{equation*}
whence
\begin{eqnarray}
\left| F_{z,\theta+h}\left(r\right)-F_{z,\theta}\left(r\right)\right| &\leq& P\left(\left|\theta^\top X - r\right| \leq \left|h^\top X\right|\mid Z=z\right)\label{eq:tech1-1}\\
&\leq& P\left(\left|\theta^\top X - r\right| \leq \kappa_4 \|h\|_2\mid Z=z\right)\label{eq:tech1-2}\\
&\leq& \kappa_4\kappa_6\|h\|_2,\label{eq:tech1-3}
\end{eqnarray}
where (\ref{eq:tech1-1}) is due to Jensen's inequality and (\ref{eq:tech1-2})-(\ref{eq:tech1-3}) use Assumptions \ref{a1}(iii) and \ref{a1}(v) respectively. The desired bound then follows because (\ref{eq:tech1-3}) does not depend on $r$.

\end{document}